\documentclass[12pt]{article}
\usepackage{amssymb}
\usepackage{graphicx}
\usepackage{amsmath}

\input{tcilatex}

\begin{document}

\begin{center}
{\Large Subsample Least Squares Estimator for Heterogeneous Effects of
Multiple Treatments with Any Outcome Variable\medskip }

(September 13, 2023)\medskip 

\begin{tabular}{c}
Myoung-jae Lee \\ 
Department of Economics \\ 
Korea University \\ 
Seoul 02841, Korea \\ 
myoungjae@korea.ac.kr%
\end{tabular}%
\bigskip \bigskip \bigskip \bigskip \bigskip \bigskip \bigskip 
\end{center}

For multiple treatments $D=0,1,...,J$, covariates $X$ and outcome $Y$, the
ordinary least squares estimator (OLS)\ of $Y$ on $(D_{1},...,D_{J},X)$ is
widely applied to a constant-effect linear model, where $D_{j}$ is the dummy
variable for $D=j$. However, the treatment effects are almost always $X$%
-heterogeneous in reality, or $Y$ is noncontinuous, to invalidate such a
linear model. The blind hope\ of practitioners is that the OLS
\textquotedblleft somehow\textquotedblright\ estimates a sensible average of
the unknown $X$-heterogeneous effects. This paper shows that, unfortunately,
the OLS\ is inconsistent unless all treatment effects are constant, because
the estimand of the $D_{d}$-slope involves the $X$-heterogeneous effects of
all treatments, not just $D_{d}$. One way to overcome this \textquotedblleft
contamination\textquotedblright\ problem is the OLS of $Y$ on $%
D_{d}-E(D_{d}|X,\ D=0,d)$ using only the subsample $D=0,d$, and this paper
proposes a modified version of the subsample OLS that is robust to
misspecifications of $E(D_{d}|X,\ D=0,d)$. The robustified subsample OLS is
proven to be consistent for an \textquotedblleft overlap
weight\textquotedblright\ average of the $X$-heterogeneous effect of $D_{d}$
for any form of $Y$ (continuous, binary, count, ...).\bigskip \bigskip
\bigskip \bigskip 

\textbf{Running Head}: How to find multiple treatment effects with
OLS.\medskip

\medskip

\textbf{Compliance with Ethical Standards and Conflict of Interest:}\ no
human or animal subject is involved in this research, and there is no
conflict of interest to disclose.\pagebreak

\section{Introduction}

\qquad With a binary treatment $D$, covariates $X$ and an outcome $Y$, the
ordinary least squares estimator (OLS) of $Y$ on $(D,X)$ for a
constant-effect linear model $Y=\beta _{d}D+\beta _{x}^{\prime }X+V$ is
almost ubiquitously used in practice, where the $\beta $'s are parameters
and $V$ is an error term. However, in reality, the treatment effect is never
a constant, but an unknown function $\mu (X)$ of $X$, which makes the linear
model invalid. Also, if $Y$ is noncontinuous, then the linear model is
invalid in general. Practitioners using the OLS have a \textquotedblleft
blind hope\textquotedblright\ that the OLS $D$-slope is somehow consistent
for $E\{\mu (X)\}$, or for a sensible weighted average $E\{\omega (X)\mu
(X)\}$ with a weight $\omega (X)$. This seems to be based on a random effect
\textquotedblleft legacy\textquotedblright : when the effect of $D$ is
random, say $\beta _{di}=E(\beta _{di})+error$ that varies across the
subjects $i=1,...,N$, the OLS\ estimates $\beta _{d}\equiv E(\beta _{di})$.

\qquad Angrist (1998) and Angrist and Pischke (2008), among others, have
shown that if $E(D|X)$ is equal to the linear projection of $D|X$, i.e., if%
\begin{equation*}
E(D|X)=L(D|X)\ \ \ \ \ \text{where \ \ \ \ }L(D|X)\equiv E(DX^{\prime
})\{E(XX^{\prime })\}^{-1}X,
\end{equation*}%
then the OLS\ $D$-slope is consistent for the following \textquotedblleft
overlap weight (OW)\textquotedblright\ average of $\mu (X)$, which in fact
holds for any form of $Y$ (continuous, binary, count, ...):%
\begin{equation*}
E\{\frac{\pi _{X}(1-\pi _{X})}{E\{\pi _{X}(1-\pi _{X})\}}\mu (X)\}\text{ \ \
\ \ where \ \ \ \ }\pi _{X}\equiv E(D|X).
\end{equation*}%
However, the condition $E(D|X)=L(D|X)$ is violated if there is any
continuous covariate. Lee et al. (2023) showed that, if $E(D|X)\neq L(D|X)$,
then the OLS is inconsistent because its estimand can be non-zero even when $%
\mu (X)=0$ for all $X$.

\qquad Now, consider a multiple (or multi-valued) treatment $D=0,1,...,J$
which appears often: types (multinomial), ranks (ordinal), counts
(cardinal), etc. Defining $1[A]\equiv 1$ if $A$ holds and $0$ otherwise, the
popular practice is creating dummy variables $D_{j}\equiv 1[D=j]$, $%
j=1,...,J $, to apply the OLS of $Y$ on $(D_{1},...,D_{J},X)$. Again, the
OLS with $(D_{1},...,D_{J},X)$ is based on the premise that the treatment
effects are constants. So the same question arises:\ what does the OLS\
estimate, when the effect of $D_{d}$ is $X$-heterogeneous, say $\mu _{d}(X)$%
? \ The blind hope\ is again that the OLS $D_{d}$-slope is consistent for $%
E\{\mu _{d}(X)\} $ or a sensible weighted average of $\mu _{d}(X)$.
Unfortunately, this paper shows that the OLS estimand of $D_{d}$ is not a
(weighted) average of $\mu _{d}(X)$, but a sum of weighted averages of 
\textit{all} $\mu _{j}(X)$, $j=1,...,J$, even when $E(D_{j}|X)=L(D_{j}|X)$
for all $j$. So the popular OLS of $Y$ on $(D_{1},...,D_{J},X)$ is
inconsistent; it is the wrong way to find multiple treatment effects.

\qquad \textit{The goal of this paper is to prove the inconsistency of the
usual OLS, and then propose an alternative\textquotedblleft OLS using only
the subsample }$D=0,d$ \textit{with centered variables\textquotedblright ,
which is consistent for an OW average of }$\mu _{d}(X)$\textit{\ for any
form of }$Y$.\textit{\ }OW may look strange, but there is a fast growing
literature on OW as is shown below, and OW is adopted also in the recent
machine-learning-based heterogeneous treatment effect literature (Athey et
al. 2019 and Nie and Wager 2021, among others). More details of our goals
are provided in the remainder of this section, with the full expositions
deferred to Sections 2 and 3.\bigskip

\qquad Let $(Y^{0},Y^{1},...,Y^{J})$ be the potential outcomes for $%
D=0,1,...,J$ to have $Y=\sum_{j=0}^{J}D_{j}Y^{j}$. With `$\amalg $' denoting
independence, take $E(\cdot |X,D_{j}=1)$ on $Y$:%
\begin{eqnarray}
&&E(Y|X,D_{j}=1)=E(Y^{j}|X,D_{j}=1)=E(Y^{j}|X)\ \ \ \text{(under\ }%
D_{j}\amalg Y^{j}|X\text{)}  \TCItag{1.1} \\
&&\ =E(Y^{j}-Y^{0}|X)+E(Y^{0}|X)  \notag \\
&&\ \Longrightarrow \
E(Y|X,D)=\sum_{j=1}^{J}E(Y^{j}-Y^{0}|X)D_{j}+E(Y^{0}|X);  \TCItag{1.2}
\end{eqnarray}%
the assumption `$D_{j}\amalg Y^{j}|X$' appeared in Imbens (2000).

\qquad Defining $U\equiv Y-E(Y|X,D)$ for (1.2) renders a \textquotedblleft
linear-in-$D_{j}$ representation\textquotedblright\ for any form of $Y$\
(continuous, binary,\ count, ...), as long as $Y^{j}-Y^{0}$ makes sense:%
\begin{equation}
Y=\sum_{j=1}^{J}\mu _{j}(X)D_{j}+E(Y^{0}|X)+U\text{, }\mu _{j}(X)\equiv
E(Y^{j}-Y^{0}|X),\ U\equiv Y-E(Y|X,D).  \tag{1.3}
\end{equation}%
This linear-in-$D_{j}$ representation needs no parametric assumption
whatsoever. Substituting (1.3) into the $Y$ in the usual OLS\ formula will
reveal what the OLS actually estimates.

\qquad Suppose $E(D_{j}|X)=L(D_{j}|X)$ for all $j$; this holds for
\textquotedblleft saturated\ models\textquotedblright , i.e., if $X$ is
discrete and a full set of dummies are used for all values of $X$. A
surprising finding of this paper is that, when $J=2$ (i.e., 3 categories),
the OLS $D_{1}$-slope is consistent for%
\begin{equation}
E\{\omega _{11}(X)\mu _{1}(X)\}+E\{\omega _{12}(X)\mu _{2}(X)\}\text{ \ \
where }E\{\omega _{11}(X)\}=1,\text{ }E\{\omega _{12}(X)\}=0.  \tag{1.4}
\end{equation}%
If $\mu _{1}(X)$ and $\mu _{2}(X)$ are constants, say $\beta _{1}$ and $%
\beta _{2}$, then (1.4) becomes $\beta _{1}E\{\omega _{11}(X)\}+\beta
_{2}E\{\omega _{12}(X)\}=\beta _{1}$: the usual OLS $D_{1}$-slope is
consistent if all treatment effects are constant. Otherwise, $E\{\omega
_{12}(X)\mu _{2}(X)\}\neq 0$ in general, making the $D_{1}$-slope of the OLS
inconsistent because its estimand involves $\mu _{2}(X)$.

\qquad Still under $E(D_{j}|X)=L(D_{j}|X)$ for all $j$, generalizing (1.4)
for $J=3$ renders%
\begin{equation}
\sum_{j=1}^{J}E\{\omega _{1j}(X)\mu _{j}(X)\},\text{ \ \ }E\{\omega
_{11}(X)\}=1,\text{ }E\{\omega _{12}(X)\}=\cdots =E\{\omega _{1J}(X)\}=0. 
\tag{1.5}
\end{equation}%
Proving (1.5) for $J\geq 4$ is involved, as it requires inverting a matrix
of dimension $4\times 4$ or higher, but (1.4) with $J=2$ and (1.5) with $J=3$
should be enough to make the point that the \textit{OLS with multiple
treatments is inconsistent when the effects are heterogeneous}.

\qquad One might think that (1.4) and (1.5) are \textquotedblleft
artifacts\textquotedblright\ due to $E(D_{j}|X)=L(D_{j}|X)$, but that is not
the case. For binary $D$ and any $Y$, Lee et al. (2023) showed that the
estimand of the $D$-slope in the OLS of $Y$ on $(D,X)$ is a weighted average
of $\mu (X)\equiv E(Y^{1}-Y^{0}|X)$ plus a bias, and the OLS $D$-slope is
inconsistent because its estimand is not zero even when $\mu (X)=0$ for all $%
X$; if $E(D|X)=L(D|X)$, however, then the bias is zero, and the OLS $D$%
-slope estimand becomes the OW average of $\mu (X)$. To proceed analogously,
we imposed the condition $E(D_{j}|X)=L(D_{j}|X)$ for multiple treatments in
(1.4) and (1.5), which was just to make sense of the OLS of $Y$ on $%
(D_{1},...,D_{J},X)$, because without the condition, the OLS estimand would
be some weighted averages of the $\mu _{j}(X)$'s plus biases.

\qquad The best way to overcome the OLS inconsistency problem with multiple
treatments is using only the subsample $D=0,d$, as this would block the
\textquotedblleft contaminations\textquotedblright\ from the other
categories. However, doing the subsample OLS of $Y$ on $(D_{d},X)$ yields a
sensible weighted average of $\mu _{d}(X)$ still only under $%
E(D_{d}|X)=L(D_{d}|X)$. To drop this restrictive condition, we thus propose
the \textit{\textquotedblleft subsample OLS\textquotedblright\ of }$Y$%
\textit{\ on }$D_{d}-E(D_{d}|X,\ D=0,d)$, which turns out to be consistent
for an OW average\ of $\mu _{d}(X)$.

\qquad Although $E(D_{d}|X,\ D=0,d)$ can be estimated nonparametrically, we
parametrize $E(D_{d}|X,\ D=0,d)$ to make the subsample OLS practical. Then,
one remaining concern is that there might be misspecifications in
parametrizing $E(D_{d}|X,\ D=0,d)$. To dissipate the concern, we thus modify
the subsample OLS into the OLS of $Y-E(Y|X,\ D=0,d)$ on $D_{d}-E(D_{d}|X,\
D=0,d)$, which turns out to be \textquotedblleft
double-debiasing\textquotedblright\ (DDB; Chernozhukov et al. 2018, 2022),
being robust to misspecified $E(Y|X,\ D=0,d)$ and $E(D_{d}|X,\ D=0,d)$.

\qquad For the DDB subsample OLS, $E(Y|X,\ D=0,d)$ and $E(D_{d}|X,\ D=0,d)$
are to be estimated by a machine learning method, and then sample splitting
or cross-fitting has to be done. Since this is cumbersome and $X$ can be
high-dimensional, as a compromise between practicality and robustness, we
propose the \textit{robustified subsample OLS of }$Y-E(Y|\pi _{X}^{0},\pi
_{X}^{d},\ D=0,d)$\textit{\ on }$D_{d}-E(D_{d}|\pi _{X}^{0},\pi _{X}^{d},\
D=0,d)$\textit{\ where }$\pi _{X}^{d}\equiv E(D_{d}|X,\ D=0,d)$. Although
this is not exactly DDB, it can be close to DDB, as will be argued later.

\qquad In short, the goal of this paper is proving (1.4) and (1.5), and then
explaining the (robustified) subsample OLS. These matter greatly, given the
nearly ubiquitous application of the usual OLS\ of $Y$ on $%
(D_{1},...,D_{J},X)$ for multiple treatments, despite that the treatment
effects are heterogeneous in reality to make the usual OLS\ inconsistent.

\qquad In the remainder of this paper, Section 2 establishes (1.4), with the
proof for (1.5) with $J=3$ relegated to an online appendix. Section 3
details the (robustified) subsample OLS. Section 4 presents a simulation
demonstration of (1.4) and the (robustified) subsample OLS. Finally, Section
5 concludes this paper. Most proofs are in the online appendix.

\section{Multiple Treatment with Three Categories}

\qquad Define the linear projection $L(D_{d}|X)$ of $D_{d}$ on $X$:%
\begin{equation}
\lambda _{dX}\equiv L(D_{d}|X)\equiv E(D_{d}X^{\prime })\{E(XX^{\prime
})\}^{-1}X.  \tag{2.1}
\end{equation}%
With $J=2$, the usual OLS of $Y$ on $(D_{1},D_{2},X)$ assumes a linear model
such as $Y=\beta _{1}D_{1}+\beta _{2}D_{2}+\beta _{x}^{\prime }X+V$, where $%
(\beta _{1},\beta _{2},\beta _{x})$ are parameters and $V$ is an error term
with $E(V|D,X)=0$. The estimands of the OLS slopes of $D_{1}$ and $D_{2}$
can be found by the OLS applied to the \textquotedblleft $X$-partialled out
version\textquotedblright : take $L(\cdot |X)$ on the linear model to get $%
L(Y|X)=\beta _{1}\lambda _{1X}+\beta _{2}\lambda _{2X}+\beta _{x}^{\prime }X$%
, and then subtract this from the linear model to obtain%
\begin{equation}
Y-L(Y|X)=\beta _{1}(D_{1}-\lambda _{1X})+\beta _{2}(D_{2}-\lambda _{2X})+V. 
\tag{2.2}
\end{equation}

\qquad The estimands of the OLS of $Y-L(Y|X)$ on $D_{1}-\lambda _{1X}$ and $%
D_{2}-\lambda _{2X}$ involve $L(\cdot |X)$ appearing inside of $E(\cdot )$,
which makes interpreting the estimands difficult. Hence, as in Angrist
(1998) and Angrist and Pischke (2009), assume for a while that, as was
invoked just before (1.4),%
\begin{equation}
\pi _{jX}=\lambda _{jX}\text{ for all }j\text{\ \ \ \ \ \ \ where \ \ }\pi
_{jX}\equiv E(D_{j}|X),\text{ \ }\lambda _{jX}\equiv L(D_{j}|X).  \tag{2.3}
\end{equation}%
With $Cov$ standing for covariance, define%
\begin{equation}
C_{jd}(X)\equiv Cov(D_{j},D_{d}|X)\text{ \ \ \ \ and \ \ \ \ \ }C_{jd}\equiv
E\{C_{jd}(X)\}\text{.}  \tag{2.4}
\end{equation}%
Theorem 1 next for 3-category $D$ allows any form of $Y$, as long as $%
Y^{j}-Y^{0}$ makes sense.\bigskip

\textbf{THEOREM 1. }\textit{For }$D=0,1,2$, \textit{under }$%
C_{11}C_{22}-C_{12}^{2}\neq 0$\textit{,} $D_{j}\amalg Y^{j}|X$\textit{\ and }%
$\pi _{jX}=\lambda _{jX}$\textit{\ for all }$j$\textit{, the slopes of }$%
D_{1}$\textit{\ and }$D_{2}$\textit{\ in the OLS of }$Y$\textit{\ on }$%
(D_{1},D_{2},X)$\textit{\ are consistent for}%
\begin{eqnarray}
&&E\{\omega _{11}(X)\mu _{1}(X)+\omega _{12}(X)\mu _{2}(X)\},\ \ \ \omega
_{1j}(X)\equiv \frac{C_{22}C_{1j}(X)-C_{12}C_{2j}(X)}{C_{11}C_{22}-C_{12}^{2}%
},  \TCItag{2.5} \\
&&E\{\omega _{22}(X)\mu _{2}(X)+\omega _{21}(X)\mu _{1}(X)\},\ \ \ \omega
_{2j}(X)\equiv \frac{C_{11}C_{2j}(X)-C_{21}C_{1j}(X)}{C_{11}C_{22}-C_{12}^{2}%
}  \notag \\
&&\text{\textit{where} \ \ }E\{\omega _{11}(X)\}=E\{\omega _{22}(X)\}=1\text{%
,\ \ }E\{\omega _{12}(X)\}=E\{\omega _{21}(X)\}=0.  \TCItag{2.6}
\end{eqnarray}%
\textit{Consequently, the OLS\ }$D_{1}$\textit{-slope is inconsistent, with
its estimand \textquotedblleft contaminated\textquotedblright\ by }$%
E\{\omega _{12}(X)\mu _{2}(X)\}$\textit{\ that is not necessarily }$0$%
\textit{\ despite }$E\{\omega _{12}(X)\}=0$\textit{; if }$\mu _{j}(X)=\beta
_{j}$\textit{\ (a constant), then the }$D_{1}$\textit{-slope is consistent
for }$\beta _{1}$\textit{. Analogous statements hold for }$D_{2}$\textit{.}%
\bigskip

\qquad The online appendix extends (2.5) to four categories to obtain (1.5)
with $J=3$, where the result analogous to (2.6) holds: the weights with the
same two subscripts have the expected value $1$, whereas all the other
weights have the expected values $0$. Although we do not have a proof for $%
J\geq 4$, the cases $J=2,3$ are already enough to make the point that the
(widely practiced) usual OLS\ of $Y$ on $(D_{1},...,D_{J},X)$ should not be
used.\bigskip

\qquad \textbf{Remark 1. }(2.3) turns (2.2) into $Y-L(Y|X)=\beta
_{1}(D_{1}-\pi _{1X})+\beta _{2}(D_{2}-\pi _{2X})+V$. $L(Y|X)$ can be
dropped, as $L(Y|X)$ is orthogonal to $(D_{1}-\pi _{1X},D_{2}-\pi _{2X})$.
Then the OLS is the same as the Robinson (1988) approach to $%
Y=\sum_{j=1}^{J}\beta _{j}D_{j}+m(X)+V$ for an unknown $m(X)$; i.e., under
(2.3), the $X$-part is allowed to be an unknown function of $X$.\textbf{%
\bigskip }

\qquad \textbf{Remark 2. }The best way to avoid the usual-OLS\ inconsistency
in Theorem 1 would be using the subsample $D=0,d$ to find the effect of $D=d$
relative to $D=0$. As in Robins et al. (1992) and Lee (2018, 2021), we can
do the \textquotedblleft subsample OLS\textquotedblright\ of $Y$ on $%
D_{d}-E(D_{d}|X,\ D=0,d)$. The relevant heterogeneous effect remains the
same, because%
\begin{equation}
E(Y^{d}-Y^{0}|X,D=0,d)=E(Y^{d}-Y^{0}|X)=\mu _{d}(X)\text{ if\ }D_{j}\amalg
(Y^{0},Y^{d})|X\text{ for }j=0,d;  \tag{2.7}
\end{equation}%
the condition here is of stronger type than $D_{j}\amalg Y^{j}|X$ in (1.1)
that was for (1.3). The next section shows that this subsample OLS\ is
consistent for an OW average of $\mu _{d}(X)$:%
\begin{equation}
E\{\omega _{ow}^{d}(X)\mu _{d}(X)\},\ \ \ \omega _{ow}^{d}(X)\equiv \frac{%
\pi _{X}^{d}(1-\pi _{X}^{d})P(D=0,d|X)}{E\{\pi _{X}^{d}(1-\pi
_{X}^{d})P(D=0,d|X)\}},\ \pi _{X}^{d}\equiv E(D_{d}|X,\ D=0,d).  \tag{2.8}
\end{equation}

\qquad \textbf{Remark 3. }As was noted already, in the OLS\ for $Y=\beta
_{d}D+\beta _{x}^{\prime }X+V$ where $D$ is binary and the true effect of $D$
is $\mu (X)\equiv E(Y^{1}-Y^{0}|X)$, if $E(D|X)=L(D|X)$, then the OLS $D$%
-slope is consistent for $E\{\omega (X)\mu (X)\}$ where $\omega (X)\equiv
\pi _{X}(1-\pi _{X})/E\{\pi _{X}(1-\pi _{X})\}$ and $\pi _{X}\equiv E(D|X)$.
This is a special case of (2.5) with $C_{12}=0$:%
\begin{equation}
\omega _{11}(X)\equiv \frac{C_{22}C_{11}(X)-C_{12}C_{21}(X)}{%
C_{11}C_{22}-C_{12}^{2}}=\frac{C_{11}(X)}{C_{11}}=\frac{\pi _{X}(1-\pi _{X})%
}{E\{\pi _{X}(1-\pi _{X})\}}.  \tag{2.9}
\end{equation}

\qquad \textbf{Remark 4. }Regarding the OW $\omega (X)$ just above for
binary $D$, $\pi _{X}(1-\pi _{X})$ reaches its maximum at $\pi _{X}=0.5$,
and minimum at $\pi _{X}=0,1$. Considering matching with the propensity
score (PS) $\pi _{X}$ for $E(Y^{1}-Y^{0})=E\{E(Y^{1}-Y^{0}|\pi _{X})\}$,
since the subjects with $\pi _{X}\simeq 0.5$ overlap well with the opposite
group whereas\textbf{\ }those with $\pi _{X}\simeq 0,1$ do not, the name
`OW' (Li et al. 2018) is appropriate. PS matching avoids poor overlap by
removing subjects with $\pi _{X}\simeq 0,1$, which amounts to targeting for $%
E\{\omega _{uni}(X)E(Y^{1}-Y^{0}|\pi _{X})\}$, where $\omega _{uni}(X)$\ is
a step-shaped \textquotedblleft uniform weight\textquotedblright\ equal to $%
0 $ for $\pi _{X}\simeq 0,1$\ and a positive constant otherwise. The OW $%
\omega (X)$ can be viewed as a smoothed version of $\omega _{uni}(X)$%
.\bigskip

\qquad \textbf{Remark 5. }Further on the OW $\omega (X)$, using $E\{\omega
(X)\mu (X)\}$ instead of $E\{\mu (X)\}$ accords many benefits: stabilizing
`inverse probability weighting' estimators (Li and Green 2013), making
`regression adjustment' estimators robust to misspecified outcome regression
models (Vansteelandt and Daniel 2014), and so on;\ see Choi and Lee (2023)
for more advantages of OW. OW is pervasive in the recent literature:\ in
addition to the studies already mentioned, see Mao et al. (2018,\ 2019), Li
and Li (2019), Li et al. (2019), Mao and Li (2020), Thomas et al. (2020),
and Cheng et al. (2022).

\section{Subsample OLS with Propensity Score Residual}

$\qquad $Our main finding has been that, when the effects are heterogeneous,
the usual OLS\ of $Y$ on $(D_{1},...,D_{J},X)$ is inconsistent in general
due to \textquotedblleft contaminations\textquotedblright\ from
other-category effects. For example, we would hope that the $D_{d}$-slope in
the OLS is consistent for a weighted average of $\mu _{d}(X)$, but its
estimand involves $\mu _{j}(X)$, $j\neq d$. The only way to rule out such
contaminations would be using the subsample $D=0,d$. This section examines
the subsample OLS\ mentioned just before (2.7) along with its robustified
version, and then presents specific estimators for ordinal and multinomial $%
D $.

\subsection{Subsample-OLS Estimand}

\qquad The subsample OLS of $Y$ on $(D_{d},X)$ still requires a restrictive
condition as in (2.3):%
\begin{equation}
E(D_{d}|X,D^{0d}=1)=L(D_{d}|X,D^{0d}=1)\text{ \ \ where\ \ \ }D^{0d}\equiv
1[D=0,d]=D_{0}+D_{d}.  \tag{3.1}
\end{equation}%
Recalling $\pi _{X}^{d}$ in (2.8), if $D_{d}-\pi _{X}^{d}$ were used from
the outset in the subsample OLS, then the restrictive condition (3.1) would
not be necessary. Hence, we proposed the subsample OLS\ of $Y$ on $D_{d}-\pi
_{X}^{d}$ in Remark 2, which is consistent for $E\{\omega _{ow}^{d}(X)\mu
_{d}(X)\}$ in (2.8).\bigskip

\textbf{THEOREM\ 2}. \textit{Under }$E\{D^{0d}(D_{d}-\pi _{X}^{d})^{2}\}>0$, 
$D_{j}\amalg (Y^{0},Y^{d})|X$\textit{\ for }$j=0,d$ \textit{in (2.7), and }$%
P(D^{0d}=1|X)>0$\textit{\ for all }$X$\textit{, the OLS }$\hat{\beta}%
_{d}^{0} $ \textit{of }$D^{0d}Y$\textit{\ on }$D^{0d}(D_{d}-\pi _{X}^{d})$%
\textit{\ is consistent for }$\beta _{d}\equiv E\{\omega _{ow}^{d}(X)\mu
_{d}(X)\}$\textit{, which holds for any }$Y$\textit{\ as long as }$%
Y^{d}-Y^{0}$\textit{\ makes sense.}\bigskip

\qquad Although Theorem 2 assumes $P(D^{0d}=1|X=x)>0$ for all $x$, if $%
P(D^{0d}=1|X)>0$ only for a set of values $x$, say $\tilde{x}$, then only
the observations with $X\in \tilde{x}$ should be used, and Theorem 2 holds
with $E\{\omega _{ow}^{d}(X)\mu _{d}(X)|X\in \tilde{x}\}$ under $%
E\{D^{0d}(D_{d}-\pi _{X}^{d})^{2}|X\in \tilde{x}\}>0$. Note that, due to $%
D^{0d}D_{d}=(D_{0}+D_{d})D_{d}=D_{d}$,%
\begin{eqnarray}
&&\ \pi _{X}^{d}\equiv E(D_{d}|X,\ D^{0d}=1)=\frac{E(D_{d}|X)}{E(D^{0d}|X)}=%
\frac{\pi _{dX}}{\pi _{0X}+\pi _{dX}};  \TCItag{3.2} \\
&&\ \omega _{ow}^{d}(X)\equiv \frac{\pi _{X}^{d}(1-\pi _{X}^{d})E(D^{0d}|X)}{%
E\{\pi _{X}^{d}(1-\pi _{X}^{d})E(D^{0d}|X)\}}=\frac{%
E(D_{0}|X)E(D_{d}|X)/E(D^{0d}|X)}{E\{E(D_{0}|X)E(D_{d}|X)/E(D^{0d}|X)\}}. 
\notag
\end{eqnarray}

\qquad For the actual implementation, we apply the subsample OLS\ of $%
D^{0d}(Y-G_{X})$ on $D^{0d}(D_{d}-\pi _{X}^{d})$ for a function $G_{X}$ of $%
X $. Using the \textquotedblleft centered\textquotedblright\ variables $%
(Y-G_{X},\ D_{d}-\pi _{X}^{d})$ instead of $(Y,\ D_{d}-\pi _{X}^{d})$ offers
advantages similar to those of DDB in case $\pi _{X}^{d}$ is misspecified.
For DDB, we set $G_{X}=E(Y|X,D^{0d}=1)$, which is to be estimated by a
machine learning method along with \textquotedblleft
cross-fitting\textquotedblright . However, since estimating $E(Y|X,D^{0d}=1)$
can be involved, we search for a simpler $G_{X}$ next. Note that $G_{X}$ is
irrelevant for the subsample OLS estimand $\beta _{d}$ in Theorem 2, because%
\begin{eqnarray}
&&E\{G_{X}D^{0d}(D_{d}-\pi _{X}^{d})\}=E[G_{X}\cdot E\{D^{0d}(D_{d}-\pi
_{X}^{d})|X\}]  \notag \\
&=&E[G_{X}\cdot \{E(D_{d}|X)-E(D^{0d}|X)\pi _{X}^{d}\}]=0\ \ \ \text{for any 
}G_{X}\text{ \ \ (using (3.2)).}  \TCItag{3.3}
\end{eqnarray}

\subsection{Robustified Subsample OLS}

\qquad Call an OLS\ using a centered treatment variable such as $D_{d}-\pi
_{X}^{d}$ \textquotedblleft OLS$_{psr}$\textquotedblright\ (Lee 2018), where
`psr' stands for `PS\ residual'. In order to motivate robustified subsample
OLS$_{psr}$, we intuitively explain first why $G_{X}=E(Y|X,D^{0d}=1)$
renders DDB, omitting regularity conditions. Then we show that $%
G_{X}=E(Y|\pi _{0X},\pi _{dX},D^{0d}=1)$ works similarly, although it falls
short of DDB.

\qquad The proof for Theorem 2 in the online appendix reveals in (A.5) that
taking $E(\cdot |X)$ on $D^{0d}Y=D^{0d}\{\sum_{j=1}^{J}\mu
_{j}(X)D_{j}+E(Y^{0}|X)+U\}$ leads to a \textquotedblleft projection
residual model\textquotedblright :%
\begin{equation}
D^{0d}\{Y-E(Y|X,D^{0d}=1)\}=\mu _{d}(X)\cdot D^{0d}(D_{d}-\pi
_{X}^{d})+D^{0d}U.  \tag{3.4}
\end{equation}%
The moment condition for $\beta _{d}$ in this projection residual model is%
\begin{equation*}
E[\ D^{0d}\{Y-E(Y|X,D^{0d}=1)-\beta _{d}(D_{d}-\pi _{X}^{d})\}\cdot
(D_{d}-\pi _{X}^{d})\ ]=0,
\end{equation*}%
because solving this for $\beta _{d}$ renders $\beta _{d}$ in Theorem 2.
Note that $E(Y|X,D^{0d}=1)$ is irrelevant for $\beta _{d}$, although it is
relevant for DDB, which is shown next.

\qquad To see that $G_{X}=E(Y|X,D^{0d}=1)$ leads to DDB, replace $%
E(Y|X,D^{0d}=1)$ with $E(Y|X,D^{0d}=1)+ah_{X}$ in the moment above, and
replace $\pi _{X}^{d}$ with $\pi _{X}^{d}+cp_{X}$, where $(a,c)$ are
constants and $h_{X}$ and $p_{X}$ are functions of $X$:%
\begin{equation*}
E[\ D^{0d}\{Y-E(Y|X,D^{0d}=1)-ah_{X}-\beta _{d}(D_{d}-\pi
_{X}^{d}-cp_{X})\}\cdot (D_{d}-\pi _{X}^{d}-cp_{X})\ ].
\end{equation*}%
Differentiate this with respect to (wrt) $(a,c)$ and evaluate the
derivatives at $(0,0)$:%
\begin{eqnarray*}
&&\ -E\{h_{X}D^{0d}(D_{d}-\pi _{X}^{d})\}=0\text{ \ \ \ \ due to (3.3) with }%
G_{X}\text{ replaced by }h_{X}; \\
&&\ \beta _{d}E\{p_{X}D^{0d}(D_{d}-\pi
_{X}^{d})\}-E[p_{X}D^{0d}\{Y-E(Y|X,D^{0d}=1)-\beta _{d}(D_{d}-\pi
_{X}^{d})\}]=0\text{.}
\end{eqnarray*}%
Since these derivatives are zero, the subsample OLS$_{psr}$\ using $%
G_{X}=E(Y|X,D^{0d}=1)$ is DDB (Chernozhukov et al. 2018, 2022).

\qquad Despite the DDB property, to avoid the possible high-dimensional
estimation problem in $G_{X}=E(Y|X,D^{0d}=1)$, we propose the robustified
subsample OLS$_{psr}$ with $G_{X}=E(Y|\pi _{0X},\pi _{dX},D^{0d}=1)$, which
appears in the projection residual model (3.4) when the projection is done
with the two-dimensional $(\pi _{0X},\pi _{dX})$ instead of $X$.\bigskip

\textbf{THEOREM 3}. \textit{Under the same conditions as in Theorem 2, the
projection residual model using }$(\pi _{0X},\pi _{dX})$\textit{\ instead of 
}$X$\textit{\ renders }%
\begin{eqnarray*}
&&D^{0d}\{Y-E(Y|\pi _{X}^{0d},D^{0d}=1)\}=\mu _{d}(\pi _{X}^{0d})\cdot
D^{0d}(D_{d}-\pi _{X}^{d})+D^{0d}U^{\prime }\ \ \ \ \ \text{where} \\
&&\pi _{X}^{0d}\equiv (\pi _{0X},\pi _{dX})^{\prime },\text{\ \ \ }\mu
_{d}(\pi _{X}^{0d})\equiv E(Y^{d}-Y^{0}|\pi _{X}^{0d}),\ \ \ U^{\prime
}\equiv Y-E(Y|\pi _{X}^{0d},D^{0d}=1).
\end{eqnarray*}%
\textit{The robustified subsample OLS}$_{psr}$\textit{\ }$\hat{\beta}%
_{d}^{\pi }$ \textit{of }$D^{0d}\{Y-E(Y|\pi _{X}^{0d},D^{0d}=1)\}$ \textit{on%
} $D^{0d}(D_{d}-\pi _{X}^{d})\ $\textit{is consistent for the same }$\beta
_{d}$\textit{\ in Theorem 2.}\bigskip

\qquad Since $\omega _{ow}^{d}(X)$ in (2.8) depends on $X$ only through $\pi
_{X}^{0d}$---$P(D=0,d|X)$ can be written as $P(D=0,d|\pi _{X}^{0d})$---$\mu
_{d}(X)$ in $\beta _{d}$ can be replaced by $\mu _{d}(\pi _{X}^{0d})$, and $%
(D_{0},D_{d})\amalg (Y^{0},Y^{d})|X$ can be weakened to $(D_{0},D_{d})\amalg
(Y^{0},Y^{d})|\pi _{X}^{0d}$, as the proof for Theorem 3 in the online
appendix reveals. This explains why the same $\beta _{d}$ is the estimand in
Theorem 3. The robustified subsample OLS$_{psr}$ in Theorem 3 is not exactly
DDB, but the online appendix shows that to the extent that the
misspecification direction of $\pi _{X}^{0d}$ is well approximated by a
power function of $\pi _{X}^{0d}$, the robustified subsample OLS$_{psr}$ is
close to being DDB.

\qquad We will try both $E(Y|X,D^{0d}=1)$ and $E(Y|\pi _{X}^{0d},D^{0d}=1)$
for $G_{X}$ in our simulation study, where both will be seen to perform
comparably, but surprisingly, $E(Y|\pi _{X}^{0d},D^{0d}=1)$ edges out $%
E(Y|X,D^{0d}=1)$ in our simulation design. For binary $D$, the performance
of OLS$_{psr}$ was one of the best in the extensive simulation study of Lee
and Lee (2022).

\qquad Many studies use machine learning methods to estimate the
heterogeneous effect $\mu (x)\equiv E(Y^{1}-Y^{0}|X=x)$ for binary $D$.
E.g., Athey et al. (2019) adopted `causal forest'\ to find a localizing
weight $w_{i}(x)$, and estimated the OW average effect around $x$ with%
\begin{equation}
\frac{\sum_{i}w_{i}(x)\{Y_{i}-E(Y|X_{i})\}(D_{i}-\pi _{X_{i}})}{%
\sum_{i}w_{i}(x)(D_{i}-\pi _{X_{i}})^{2}},\ \ \ w_{i}(x)\equiv \frac{1}{B}%
\sum_{b=1}^{B}\frac{1[X_{i}\in L_{b}(x)],\ i\in S_{b}]}{\#\{j:X_{j}\in
L_{b}(x),\ j\in S_{b}\}}  \tag{3.5}
\end{equation}%
where $L_{b}(x)$ is the `leaf'\ to which $x$ belongs, $S_{b}$ is a subsample
to estimate the causal tree $b$, and $\#\{\cdot \}$ denotes the cardinality
of $\{\cdot \}$; $E(Y|X_{i})$ and $\pi _{X_{i}}$ are estimated by
\textquotedblleft leave-one-out\textquotedblright\ estimators not using the $%
i$th observation. Also, Nie and Wager (2021) solved%
\begin{equation}
\sum_{i}[\{Y_{i}-\widehat{E(Y|X_{i})}-(D_{i}-\hat{\pi}_{X_{i}})\mu
(X_{i})\}^{2}]+\Lambda _{n}\{\mu (\cdot )\}  \tag{3.6}
\end{equation}%
for $\mu (\cdot )$, where `\symbol{94}' denotes estimators, and $\Lambda
_{n}\{\mu (\cdot )\}$ is a \textquotedblleft regularizer\textquotedblright\
of $\mu (\cdot )$. In view of (3.4) without $D^{0d}$, (3.5) and (3.6) are to
estimate (local to $x$) OW averages of $\mu (X)$.

\subsection{Estimators for Ordinal and Multinomial Treatments}

\qquad Consider an ordered probit model: for an error term $\varepsilon $
and thresholds $\tau _{1},\tau _{2},...,\tau _{J}$,%
\begin{eqnarray*}
&&D=\sum_{j=1}^{J}1[\tau _{j}\leq X^{\prime }\kappa +\varepsilon ],\ \ \
\varepsilon \sim N(0,\sigma ^{2})\amalg X,\ \ \ X\text{ is }\nu \times 1; \\
&&\text{the identified parameter is \ }\alpha \equiv (\frac{\kappa _{1}-\tau
_{1}}{\sigma },\frac{\kappa _{2}}{\sigma },...,\frac{\kappa _{\nu }}{\sigma }%
,\text{\ }\frac{\tau _{2}-\tau _{1}}{\sigma },...,\frac{\tau _{J}-\tau _{1}}{%
\sigma })^{\prime }.
\end{eqnarray*}%
Setting $\sigma =1$ and $\tau _{1}=0$ for normalization, the parameters are
estimated by ordered probit. Then, with $\tau _{0}=-\infty $ and $\tau
_{J+1}=\infty $,%
\begin{equation}
P_{j}(\alpha ;X)\equiv P(D=j|X)=\Phi (\tau _{j+1}-X^{\prime }\kappa )-\Phi
(\tau _{j}-X^{\prime }\kappa )\text{, }\ \ j=0,1,...,J.  \tag{3.7}
\end{equation}

\qquad We use two types of dependent variables for the robustified subsample
OLS$_{psr}$. Using only the subsample $D^{0d}=1$, the first is\ $Y-\hat{Y}%
^{(1)}$ where $\hat{Y}^{(1)}$ is the predicted value from the OLS of $Y$ on
\textquotedblleft order-$q$\textquotedblright\ power functions of elements
of $X$, and the second is $Y-\hat{Y}^{(2)}$ where $\hat{Y}^{(2)}$ is the
predicted value $\sum_{p=0}^{q}(X^{\prime }\hat{\kappa})^{p}\hat{\gamma}%
_{p}^{d}$ from the OLS\ $\hat{\gamma}^{d}\equiv (\hat{\gamma}_{0}^{d},...,%
\hat{\gamma}_{q}^{d})^{\prime }$ of $Y$ on $\{1,X^{\prime }\hat{\kappa}%
,...,(X^{\prime }\hat{\kappa})^{q}\}$. The former is for $%
G_{X}=E(Y|X,D^{0d}=1)$, and the latter is for $G_{X}=E(Y|\pi
_{X}^{0d},D^{0d}=1)$ in Theorem 3. We will use mainly the latter to explain
the robustified subsample OLS$_{psr}$; how to adapt this for the former
should be easy to see.

\qquad Our estimator OLS$_{psr}^{q}$ \textquotedblleft of order $q$%
\textquotedblright\ with the above ordinal $D$ is%
\begin{equation*}
\text{OLS\ \ }\hat{\beta}_{d}^{q}\text{ \ of \ }D^{0d}\{Y-\sum_{p=0}^{q}(X^{%
\prime }\hat{\kappa})^{p}\hat{\gamma}_{p}^{d}\}\text{\ \ on\ \ }%
D^{0d}\{D_{d}-\frac{P_{d}(\hat{\alpha};X)}{P_{0}(\hat{\alpha};X)+P_{d}(\hat{%
\alpha};X)}\}.
\end{equation*}%
The sample moment condition satisfied by $\hat{\beta}_{d}^{q}$ is: with $a$
for $\alpha $, $k$ for $\kappa $, and $t_{j}$ for $\tau _{j}$,%
\begin{eqnarray}
&&\frac{1}{\sqrt{N}}\sum_{i}m(\hat{\beta}_{d}^{q},\hat{\alpha},\hat{\gamma}%
^{d};X_{i},Y_{i})=0;\text{ \ \ }m(b,a,g;X,Y)\text{ with }a\equiv (k^{\prime
},t_{2},...,t_{J})^{\prime }\text{ is}  \TCItag{3.8} \\
&&D^{0d}[Y-\sum_{p=0}^{q}(X^{\prime }k)^{p}g_{p}^{d}-b\{D_{d}-\frac{%
P_{d}(a;X)}{P_{0}(a;X)+P_{d}(a;X)}\}]\ \{D_{d}-\frac{P_{d}(a;X)}{%
P_{0}(a;X)+P_{d}(a;X)}\}.  \notag
\end{eqnarray}

\textbf{THEOREM\ 4}. \textit{For ordinal }$D$\textit{\ with (3.7), }$\sqrt{N}%
(\hat{\beta}_{d}^{q}-\beta _{d})\rightarrow ^{d}N(0,\Omega _{d})$\textit{\
for any }$q=0,1,...$\textit{,}%
\begin{eqnarray}
\hat{\Omega}_{d}^{q} &\equiv &(\frac{1}{N}\sum_{i}D_{i}^{0d}\hat{\varepsilon}%
_{di}^{2})^{-2}\cdot \frac{1}{N}\sum_{i}(D_{i}^{0d}\hat{V}_{di}\hat{%
\varepsilon}_{di}+\hat{L}_{d}\hat{\eta}_{i})^{2}\rightarrow ^{p}\Omega _{d},
\TCItag{3.9} \\
\hat{\varepsilon}_{di} &\equiv &D_{di}-\frac{P_{d}(\hat{\alpha};X_{i})}{%
P_{0}(\hat{\alpha};X_{i})+P_{d}(\hat{\alpha};X_{i})},\ \ \ \ \ \hat{V}%
_{di}\equiv Y_{i}-\sum_{p=0}^{q}(X_{i}^{\prime }\hat{\kappa})^{p}\hat{\gamma}%
_{p}^{d}-\hat{\beta}_{d}^{q}\hat{\varepsilon}_{di},  \notag \\
\hat{L}_{d} &\equiv &\frac{1}{N}\sum_{i}\frac{\partial m(\hat{\beta}_{d}^{q},%
\hat{\alpha},\hat{\gamma}^{d};X_{i},Y_{i})}{\partial a^{\prime }},\ \ \ \ \ 
\hat{\eta}_{i}\equiv (\frac{1}{N}\sum_{i}\hat{s}_{i}\hat{s}_{i}^{\prime
})^{-1}\hat{s}_{i},  \notag
\end{eqnarray}%
\textit{and }$\hat{s}_{i}$\textit{\ is the estimated ordered probit score
function using all observations.}\bigskip

\qquad Since the procedure is \textquotedblleft smooth\textquotedblright ,
if $\hat{\Omega}_{d}^{d}$ looks involved, then bootstrap may be used
instead. For a joint test involving $\beta _{d}$ and $\beta _{j}$, with the
asymptotic variances found as in (3.9), we need the asymptotic covariance of 
$\hat{\beta}_{d}^{q}$ and $\hat{\beta}_{j}^{q}$:%
\begin{equation*}
(\frac{1}{N}\sum_{i}D_{i}^{0d}\hat{\varepsilon}_{di}^{2})^{-1}\cdot \frac{1}{%
N}\sum_{i}(D_{i}^{0d}\hat{V}_{di}\hat{\varepsilon}_{di}+\hat{L}_{d}\hat{\eta}%
_{i})(D_{i}^{0j}\hat{V}_{ji}\hat{\varepsilon}_{ji}+\hat{L}_{j}\hat{\eta}%
_{i})\cdot (\frac{1}{N}\sum_{i}D_{i}^{0j}\hat{\varepsilon}_{ji}^{2})^{-1}.
\end{equation*}%
We can obtain $\partial m(\hat{\beta}_{d}^{q},\hat{\alpha},\hat{\gamma}%
;X_{i},Y_{i})/\partial a$ algebraically, but it would be less error-prone to
use numerical gradients as is done in our simulation study below.\bigskip

\qquad Suppose now that $D$ is multinomial with, for $d=0,1,...,J$,%
\begin{equation*}
P_{d}(\alpha ;X)\equiv P(D=d|X)=\frac{\exp (W_{d}^{\prime }\alpha )}{%
1+\sum_{j=1}^{J}\exp (W_{j}^{\prime }\alpha )}\text{ \ \ \ \ where \ \ }%
W_{0}^{\prime }\alpha \equiv 1,
\end{equation*}%
and $W_{j}$ consists of zeros and elements of $X$; see, e.g., Lee (2010,
p.231) for how to construct $W_{j}$. Here, $\alpha $ is to be estimated by
multinomial logit (MNL).

\qquad Differently from the ordinal $D$ with the single \textquotedblleft
index function\textquotedblright\ $X^{\prime }\kappa $ in (3.7), there
appear multiple index functions\ $W_{1}^{\prime }\alpha ,...,W_{J}^{\prime
}\alpha $ for the multinomial $D$. Let $\hat{\gamma}^{d}\equiv (\hat{\gamma}%
_{0}^{d},\hat{\gamma}_{1}^{d},...,\hat{\gamma}_{J}^{d})^{\prime }$ be the
OLS of $Y$ on $(1,W_{1}^{\prime }\hat{\alpha},...,W_{J}^{\prime }\hat{\alpha}%
)$ using the $D^{0d}=1$ subsample. Then we can use $Y-\sum_{j=0}^{J}(W_{j}^{%
\prime }\hat{\alpha})\hat{\gamma}_{j}^{d}$ with $W_{0}^{\prime }\hat{\alpha}%
\equiv 1$ as the robustified subsample OLS$_{psr}$ dependent variable. This
would be an order-1 approximation to $E(Y|W_{1}^{\prime }\alpha
,...,W_{J}^{\prime }\alpha )$, and an order-2 approximation would be using $%
(1,W_{1}^{\prime }\hat{\alpha},...,W_{J}^{\prime }\hat{\alpha})$ and their
second-order terms.

\qquad With multinomial $D$, the change needed for Theorem 4 is replacing $%
Y-\sum_{p=0}^{q}(X^{\prime }\hat{\kappa})^{p}\hat{\gamma}_{p}^{d}$ with a
properly centered version such as $Y-\sum_{j=0}^{J}(W_{j}^{\prime }\hat{%
\alpha})\hat{\gamma}_{j}^{d}$; also, $\hat{s}_{i}$ should be the MNL score
function. If an estimator other than ordered probit or multinomial logit is
used for ordinal or multinomial $D$, then Theorem 4 needs an appropriate
modification: a properly centered $Y$ and the score function for the
estimator should be used.

\section{Simulation Study}

\qquad To demonstrate that the usual OLS\ of $Y$ on $(D_{1},...,D_{J},X)$ is
inconsistent when the treatment effects are heterogeneous, consider a simple
model with an ordinal $D=0,1,2$:%
\begin{eqnarray}
&&P(X_{2}=0)=0.3\text{, }P(X_{2}=1)=0.7,\ \ \ \varepsilon \sim
N(0,0.5^{2})\amalg X_{2},  \notag \\
&&D=1[0\leq \kappa _{1}+\kappa _{2}X_{2}+\varepsilon ]+1[1\leq \kappa
_{1}+\kappa _{2}X_{2}+\varepsilon ],\ \ \ \ \ \kappa _{1}=0\text{,\ }\kappa
_{2}=1,  \TCItag{4.1} \\
&&Y^{0}=\beta _{1}+\beta _{2}X_{2}+U,\ \ Y^{d}-Y^{0}=dX_{2},\ \ \beta
_{1}=\beta _{2}=1,\ \ U\sim N(0,1)\amalg (\varepsilon ,X_{2}),  \notag \\
&&Y=\sum_{j=0}^{2}D_{j}Y^{j},\ \ \mu _{d}(X_{2})=d\times E(X_{2})\text{
which is }0.7\text{ for }d=1\text{ and }1.4\text{ for }d=2.  \notag
\end{eqnarray}%
Here, given $Y^{d}-Y^{0}=dX_{2}$, one might naively expect for the OLS to be
consistent for%
\begin{equation}
\{\beta _{1},\ E(X_{2}),\ 2E(X_{2}),\ \beta _{2}\}=(1,\ 0.7,\ 1.4,\ 1). 
\tag{4.2}
\end{equation}%
Let $N_{0}$ and $N_{1}$ be the sample size for $X_{2}=0$ and $X_{2}=1$,
respectively.

\qquad Estimating $C_{jd}(X_{2})\equiv Cov(D_{j},D_{d}|X_{2})$ and $%
C_{jd}\equiv E\{C_{jd}(X_{2})\}$, we have $C_{jd}=0.3C_{jd}(0)+0.7C_{jd}(1)$
where, with $\bar{D}_{j}$ denoting the sample mean of $D_{j}$,%
\begin{equation*}
C_{jd}(0)\simeq \frac{1}{N_{0}}\sum_{i:X_{2}=0}(D_{ij}-\bar{D}_{j})(D_{id}-%
\bar{D}_{d}),\text{ \ \ }C_{jd}(1)\simeq \frac{1}{N_{1}}%
\sum_{i:X_{2}=1}(D_{ij}-\bar{D}_{j})(D_{id}-\bar{D}_{d}).
\end{equation*}%
For $d=1,2$, due to $\mu _{d}(0)=E(Y^{d}-Y^{0}|X_{2}=0)=dX_{2}=0$ and $\mu
_{d}(1)=E(Y^{d}-Y^{0}|X_{2}=1)=dX_{2}=d$, the OLS estimand for $%
(D_{1},D_{2}) $ in Theorem 1 is (recall $P(X_{2}=1)=0.7$)%
\begin{eqnarray*}
E\{\omega _{11}(X_{2})\mu _{1}(X_{2})+\omega _{12}(X_{2})\mu _{2}(X_{2})\}
&=&\{\omega _{11}(1)\times 1+\omega _{12}(1)\times 2\}0.7\simeq 0.13, \\
E\{\omega _{22}(X_{2})\mu _{2}(X_{2})+\omega _{21}(X_{2})\mu _{1}(X_{2})\}
&=&\{\omega _{22}(1)\times 2+\omega _{21}(1)\times 1\}0.7\simeq 1.13,
\end{eqnarray*}%
where $0.13$ and $1.13$ come from a sample of $N=1,000,000$. For the same
sample, the OLS of $Y$ on $(1,D_{1},D_{2},X_{2})$ gave four estimates
(t-values):%
\begin{equation*}
0.91\ (358),\ \ 0.13\ (36.5),\ \ 1.12\ (265),\ \ 1.94\ (633).
\end{equation*}%
This clearly demonstrates that the OLS estimand is what Theorem 1 states,
not (4.2).\bigskip

\qquad In (4.1), we used binary $X_{2}$ to easily compute the treatment
effects and OLS estimand. Now we use a longer simulation model for our
subsample OLS$_{psr}$: with $Uni[0,2]$ denoting the uniform distribution on $%
[0,2]$, $N=1,000$ and $4,000$, and $5,000$ repetitions,%
\begin{eqnarray}
&&X_{2}\sim N(0,1),\ X_{3}\sim Uni[0,2],\ \ \varepsilon \sim N(0,1)\text{ or
standardized }\chi _{3}^{2}\amalg (X_{2},X_{3})  \notag \\
&&D^{\ast }\equiv \kappa _{1}+\kappa _{2}X_{2}+\kappa _{3}X_{3}+\varepsilon 
\text{,\ \ \ }D=1[0\leq D^{\ast }]+1[1\leq D^{\ast }],\ \ (\kappa
_{1},\kappa _{2},\kappa _{3})=(0,1,1)  \notag \\
&&Y^{0}=\beta _{1}+\beta _{2}X_{2}+\beta _{3}X_{3}+U,\ \ \
Y^{d}-Y^{0}=dX_{3},\ \ \ (\beta _{1},\beta _{2},\beta _{3})=(1,1,1), 
\TCItag{4.3} \\
&&U\sim N(0,1)\amalg (\varepsilon ,X_{2},X_{3}),\ \ \mu
_{d}(X)=E(Y^{d}-Y^{0}|X)=dE(X_{3})=1,2\text{ \ for }d=1,2.  \notag
\end{eqnarray}%
We try two distributions for $\varepsilon $:\ normal (symmetric) and $\chi
_{3}^{2}$ (asymmetric); $\varepsilon $ is then standardized for $%
E(\varepsilon )=0$ and $SD(\varepsilon )=1$ where SD stands for standard
deviation. For $\chi _{3}^{2}$ distribution, ordered probit (grossly)
misspecifies the distribution of $\varepsilon $.

\qquad Table 1 shows the average absolute bias, the SD of the $5,000$
repetitions, along with the average of the $5,000$ SD formula in Theorem 4;
also, the root mean squared error (MSE) is reported. In computing the bias,
we calculate the true effects $\mu _{1}(X)$ and $\mu _{2}(X)$ anew using the
given sample at each repetition. For OLS$_{psr}$, we present three versions,
depending on the centered dependent variable: (i) $\hat{\beta}_{d}^{0}$ with 
$Y-\bar{Y}$; (ii) $\hat{\beta}_{d}^{X}$ with $Y-\bar{Y}^{(1)}$ where $\bar{Y}%
^{(1)}$ is the predicted value from the OLS$\ $of $Y$ on $%
(1,X_{2},X_{3},X_{2}^{2},X_{3}^{2},X_{2}X_{3})$ using the subsample $%
D^{01}=1 $, which is for $G_{X}=E(Y|X,D^{0d}=1)$; and (iii) $\hat{\beta}%
_{d}^{\pi }$ with $Y-\bar{Y}^{(2)}$ where $\bar{Y}^{(2)}=\sum_{p=0}^{2}(X^{%
\prime }\hat{\kappa})^{p}\hat{\gamma}_{p}^{d}$, which is for $G_{X}=E(Y|\pi
_{X}^{0d},D^{0d}=1)$.

\qquad Table 1 presents the results in four panels: (1) $\varepsilon \sim
N(0,1)$, (2)$\ \varepsilon \sim \chi _{3}^{2}$, (3) $\varepsilon \sim N(0,1)$
with the ordered probit regression misspecified by omitting $X_{2}^{2}$ with
slope $1$, and (4) $\varepsilon \sim \chi _{3}^{2}$\ with the same
misspecification omitting $X_{2}^{2}$. In Panel 1 where `$\varepsilon \sim
N(0,1)$' is correct, all three estimators perform similarly. In Panel 2 with
the $\varepsilon $ distribution misspecified, $\hat{\beta}_{d}^{0}$ is
highly biased, whereas $\hat{\beta}_{d}^{X}$ and $\hat{\beta}_{d}^{\pi }$
are much less biased or almost unbiased; also their SD's are lower than the
SD of $\hat{\beta}_{d}^{0}$; $\hat{\beta}_{d}^{X}$ and $\hat{\beta}_{d}^{\pi
}$ perform almost the same.

\begin{center}
$%
\begin{tabular}{cccc}
\hline\hline
\multicolumn{4}{c}{Table 1: OLS$_{psr}$ Results for Ordinal $D=0,1,2$ ($%
5,000 $ Repetitions)} \\ 
&  & $N=1,000$ & $N=4,000$ \\ 
&  & \TEXTsymbol{\vert}Bias\TEXTsymbol{\vert} (SD, SD$_{asy}$) RMSE & 
\TEXTsymbol{\vert}Bias\TEXTsymbol{\vert} (SD, SD$_{asy}$) RMSE \\ \hline
\multicolumn{1}{l}{(1) $\varepsilon \sim N(0,1)$} & \multicolumn{1}{l}{$\hat{%
\beta}_{1}^{0}$} & 0.00 (0.14, 0.14) 0.14 & 0.01 (0.07, 0.07) 0.07 \\ 
\multicolumn{1}{l}{} & \multicolumn{1}{l}{$\ \ \ \hat{\beta}_{1}^{X}$} & 
0.01 (0.10, 0.10) 0.11 & 0.01 (0.05, 0.05) 0.05 \\ 
\multicolumn{1}{l}{} & \multicolumn{1}{l}{$\ \ \ \hat{\beta}_{1}^{\pi }$} & 
\textit{0.00 (0.11, 0.11) 0.11} & \textit{0.01 (0.05, 0.05) 0.06} \\ 
\multicolumn{1}{l}{} & \multicolumn{1}{l}{$\hat{\beta}_{2}^{0}$} & 0.00
(0.16, 0.17) 0.16 & 0.03 (0.08, 0.08) 0.09 \\ 
\multicolumn{1}{l}{} & \multicolumn{1}{l}{$\ \ \ \hat{\beta}_{2}^{X}$} & 
0.00 (0.14, 0.14) 0.14 & 0.03 (0.07, 0.07) 0.08 \\ 
\multicolumn{1}{l}{} & \multicolumn{1}{l}{$\ \ \ \hat{\beta}_{2}^{\pi }$} & 
\textit{0.00 (0.13, 0.13) 0.13} & \textit{0.03 (0.07, 0.07) 0.07} \\ 
\multicolumn{1}{l}{(2)$\ \varepsilon \sim \chi _{3}^{2}$} & 
\multicolumn{1}{l}{$\hat{\beta}_{1}^{0}$} & 0.46 (0.16, 0.16) 0.49 & 0.47
(0.08, 0.08) 0.48 \\ 
\multicolumn{1}{l}{} & \multicolumn{1}{l}{$\ \ \ \hat{\beta}_{1}^{X}$} & 
0.06 (0.11, 0.11) 0.13 & 0.03 (0.06, 0.06) 0.07 \\ 
\multicolumn{1}{l}{} & \multicolumn{1}{l}{$\ \ \ \hat{\beta}_{1}^{\pi }$} & 
\textit{0.05 (0.12, 0.12) 0.13} & \textit{0.03 (0.06, 0.06) 0.07} \\ 
\multicolumn{1}{l}{} & \multicolumn{1}{l}{$\hat{\beta}_{2}^{0}$} & 0.39
(0.21, 0.22) 0.44 & 0.36 (0.10, 0.11) 0.37 \\ 
\multicolumn{1}{l}{} & \multicolumn{1}{l}{$\ \ \ \hat{\beta}_{2}^{X}$} & 
0.16 (0.16, 0.16) 0.23 & 0.13 (0.08, 0.08) 0.15 \\ 
\multicolumn{1}{l}{} & \multicolumn{1}{l}{$\ \ \ \hat{\beta}_{2}^{\pi }$} & 
\textit{0.16 (0.16, 0.16) 0.22} & \textit{0.13 (0.08, 0.08) 0.15} \\ 
\multicolumn{1}{l}{(3) $\varepsilon \sim N(0,1)$} & \multicolumn{1}{l}{$\hat{%
\beta}_{1}^{0}$} & 0.00 (0.17, 0.17) 0.17 & 0.02 (0.09, 0.09) 0.09 \\ 
reg. false & \multicolumn{1}{l}{$\ \ \hat{\beta}_{1}^{X}$} & 0.08 (0.12,
0.12) 0.14 & 0.05 (0.06, 0.06) 0.08 \\ 
\multicolumn{1}{l}{} & \multicolumn{1}{l}{$\ \ \hat{\beta}_{1}^{\pi }$} & 
\textit{0.07 (0.12, 0.12) 0.14} & \textit{0.05 (0.06, 0.06) 0.08} \\ 
\multicolumn{1}{l}{} & \multicolumn{1}{l}{$\hat{\beta}_{2}^{0}$} & 0.28
(0.13, 0.15) 0.31 & 0.26 (0.07, 0.07) 0.27 \\ 
\multicolumn{1}{l}{} & \multicolumn{1}{l}{$\ \ \ \hat{\beta}_{2}^{0}$} & 
0.24 (0.12, 0.13) 0.27 & 0.21 (0.06, 0.07) 0.22 \\ 
& \multicolumn{1}{l}{$\ \ \ \hat{\beta}_{2}^{\pi }$} & \textit{0.12 (0.13,
0.13) 0.18} & \textit{0.10 (0.07, 0.06) 0.12} \\ 
\multicolumn{1}{l}{(4) $\varepsilon \sim \chi _{3}^{2}$} & 
\multicolumn{1}{l}{$\hat{\beta}_{1}^{0}$} & 0.45 (0.18, 0.17) 0.48 & 0.45
(0.09, 0.09) 0.46 \\ 
reg. false & \multicolumn{1}{l}{$\ \ \ \hat{\beta}_{1}^{X}$} & 0.21 (0.11,
0.13) 0.24 & 0.20 (0.06, 0.06) 0.21 \\ 
& \multicolumn{1}{l}{$\ \ \ \hat{\beta}_{1}^{\pi }$} & \textit{0.18 (0.13,
0.13) 0.22} & \textit{0.18 (0.06, 0.06) 0.19} \\ 
& \multicolumn{1}{l}{$\hat{\beta}_{2}^{0}$} & 0.44 (0.16, 0.18) 0.47 & 0.42
(0.08, 0.09) 0.43 \\ 
& \multicolumn{1}{l}{$\ \ \ \hat{\beta}_{2}^{X}$} & 0.48 (0.12, 0.13) 0.50 & 
0.47 (0.06, 0.06) 0.47 \\ 
\multicolumn{1}{l}{} & \multicolumn{1}{l}{$\ \ \ \hat{\beta}_{2}^{\pi }$} & 
\textit{0.32 (0.14, 0.13) 0.35} & \textit{0.31 (0.07, 0.06) 0.32} \\ \hline
\multicolumn{4}{c}{SD: simulation SD; SD$_{asy}$:\ averaged asy. SD; reg.
false: probit regression} \\ 
\multicolumn{4}{c}{misspecified; $\hat{\beta}_{d}^{0},\hat{\beta}_{d}^{X},%
\hat{\beta}_{d}^{\pi }$ with $E(Y),\ E(Y|X,D^{0d}=1),$ $E(Y|\pi
_{X}^{0d},D^{0d}=1)$} \\ \hline\hline
\end{tabular}%
$
\end{center}

\qquad In Panel 3 of Table 1 where the $\varepsilon $ distribution is
correct but the probit regression function misspecified, $\hat{\beta}%
_{d}^{\pi }$ using the ordered probit regression function for $E(Y|\pi
_{X}^{0d},D^{0d}=1)$ nearly dominates $\hat{\beta}_{d}^{X}$ using the power
function approximation for $E(Y|X,D^{0d}=1)$; the same can be said for Panel
4 where both the $\varepsilon $ distribution and the probit regression
function are misspecified. In all panels,, the SD magnitudes decline as $N$
goes up, and the actual simulation SD's agree closely with the average of
the corresponding asymptotic SD's. Overall, $\hat{\beta}_{d}^{\pi }$
performs best, followed closely by $\hat{\beta}_{d}^{X}$ and then remotely
by $\hat{\beta}_{d}^{0}$.\bigskip

\qquad Turning to multinomial $D=0,1,2$, our data generating process is%
\begin{eqnarray*}
&&X_{0},X_{1},X_{2}\sim N(0,1),\ \ \ X_{3}\sim Uni[0,2]\text{ \ \ \ (}%
X_{0},X_{1},X_{2},X_{3}\text{ are independent),} \\
&&W_{1}=(-X_{0},X_{1},0,\ X_{3},0)^{\prime }\text{, \ \ }%
W_{2}=(-X_{0},0,X_{2},\ 0,X_{3})^{\prime },\ \alpha =(1,1,1,\ 1,2)^{\prime },
\\
&&P(D=0|X)=\frac{1}{1+\sum_{j=1}^{2}\exp (W_{j}^{\prime }\alpha )},\ \ \
P(D=1|X)=\frac{\exp (W_{1}^{\prime }\alpha )}{1+\sum_{j=1}^{2}\exp
(W_{j}^{\prime }\alpha )}, \\
&&Y^{0}=\beta _{1}+\beta _{3}X_{3}+U,\ \ \ Y^{d}-Y^{0}=dX_{3},\ \ \ (\beta
_{1},\beta _{3})=(1,1), \\
&&U\sim N(0,1)\amalg (\varepsilon ,X_{0},X_{1},X_{2},X_{3}),\ \ \ \mu
_{d}(X)=dE(X_{3})=1,2\text{ \ for }d=1,2.
\end{eqnarray*}%
Here, $X_{0},X_{1},X_{2}$ are for a single alternative-varying regressor for
alternative $0,1,2$, respectively, and $X_{3}$ is an alternative-constant
regressor with different slopes across alternatives.

\qquad For (grossly) misspecified multinomial logit distribution, we set%
\begin{equation*}
P(D=0|X)=\frac{1}{1+\sum_{j=1}^{2}|W_{j}^{\prime }\alpha |},\ \ \ \ \
P(D=1|X)=\frac{|W_{1}^{\prime }\alpha |}{1+\sum_{j=1}^{2}|W_{j}^{\prime
}\alpha |}
\end{equation*}%
which is called \textquotedblleft MNabs\textquotedblright\ standing for
\textquotedblleft multinomial with absolute value\textquotedblright . For
regression misspecification, an alternative-constant regressor $e^{X_{3}}$
is erroneously omitted, as we set%
\begin{equation*}
W_{1}=(-X_{0},X_{1},0,\ X_{3},e^{X_{3}},0,0)^{\prime }\text{,\ }%
W_{2}=(-X_{0},0,X_{2},\ 0,0,X_{3},e^{X_{3}})^{\prime },\ \alpha =(1,1,1,\
1,2,1,2)^{\prime }.
\end{equation*}

\begin{center}
$%
\begin{tabular}{cccc}
\hline\hline
\multicolumn{4}{c}{Table 2: OLS$_{psr}$ Results for Multinomial $D=0,1,2$ ($%
5000$ Repetitions)} \\ 
&  & $N=1,000$ & $N=4,000$ \\ 
&  & \TEXTsymbol{\vert}Bias\TEXTsymbol{\vert} (SD, SD$_{asy}$) RMSE & 
\TEXTsymbol{\vert}Bias\TEXTsymbol{\vert} (SD, SD$_{asy}$) RMSE \\ \hline
\multicolumn{1}{l}{(1) $\varepsilon \ $MNL} & \multicolumn{1}{l}{$\hat{\beta}%
_{1}^{0}$} & 0.00 (0.23, 0.23) 0.23 & 0.00 (0.11, 0.11) 0.11 \\ 
\multicolumn{1}{l}{} & \multicolumn{1}{l}{\ \ \ $\hat{\beta}_{1}^{X}$} & 
0.01 (0.13, 0.13) 0.13 & 0.01 (0.06, 0.07) 0.06 \\ 
\multicolumn{1}{l}{} & \multicolumn{1}{l}{$\ \ \ \hat{\beta}_{1}^{\pi }$} & 
\textit{0.00 (0.15, 0.15) 0.15} & \textit{0.01 (0.07 0.08) 0.07} \\ 
\multicolumn{1}{l}{} & \multicolumn{1}{l}{$\hat{\beta}_{2}^{0}$} & 0.03
(0.21, 0.21) 0.21 & 0.04 (0.10, 0.10) 0.11 \\ 
\multicolumn{1}{l}{} & \multicolumn{1}{l}{\ \ \ $\hat{\beta}_{2}^{X}$} & 
0.01 (0.14, 0.14) 0.14 & 0.05 (0.07, 0.07) 0.09 \\ 
\multicolumn{1}{l}{} & \multicolumn{1}{l}{\ \ \ $\hat{\beta}_{2}^{\pi }$} & 
\textit{0.02 (0.16, 0.15) 0.16} & \textit{0.05 (0.08, 0.08) 0.09} \\ 
\multicolumn{1}{l}{(2)$\ \varepsilon \ $MNabs} & \multicolumn{1}{l}{$\hat{%
\beta}_{1}^{0}$} & 0.10 (0.14, 0.14) 0.17 & 0.10 (0.07, 0.07) 0.12 \\ 
\multicolumn{1}{l}{} & \multicolumn{1}{l}{\ \ $\hat{\beta}_{1}^{X}$} & 0.08
(0.09, 0.09) 0.12 & 0.06 (0.05, 0.05) 0.07 \\ 
\multicolumn{1}{l}{} & \multicolumn{1}{l}{\ \ $\hat{\beta}_{1}^{\pi }$} & 
\textit{0.02 (0.11, 0.11) 0.11} & \textit{0.01 (0.05, 0.05) 0.05} \\ 
\multicolumn{1}{l}{} & \multicolumn{1}{l}{$\hat{\beta}_{2}^{0}$} & 0.12
(0.13, 0.13) 0.18 & 0.17 (0.07, 0.06) 0.18 \\ 
\multicolumn{1}{l}{} & \multicolumn{1}{l}{\ \ $\hat{\beta}_{2}^{X}$} & 0.12
(0.10, 0.10) 0.15 & 0.05 (0.05, 0.05) 0.07 \\ 
\multicolumn{1}{l}{} & \multicolumn{1}{l}{\ \ $\hat{\beta}_{2}^{\pi }$} & 
\textit{0.00 (0.12, 0.11) 0.12} & \textit{0.05 (0.06, 0.05) 0.08} \\ 
\multicolumn{1}{l}{(3) $\varepsilon $ MNL} & \multicolumn{1}{l}{$\hat{\beta}%
_{1}^{0}$} & 0.02 (0.41, 0.71) 0.41 & 0.04 (0.19, 0.33) 0.19 \\ 
\multicolumn{1}{r}{reg. false} & \multicolumn{1}{l}{\ \ $\hat{\beta}_{1}^{X}$%
} & 0.00 (0.33, 0.31) 0.33 & 0.04 (0.16, 0.16) 0.16 \\ 
\multicolumn{1}{l}{} & \multicolumn{1}{l}{\ \ $\hat{\beta}_{1}^{\pi }$} & 
\textit{0.00 (0.34, 0.36) 0.34} & \textit{0.04 (0.16, 0.17) 0.17} \\ 
\multicolumn{1}{l}{} & \multicolumn{1}{l}{$\hat{\beta}_{2}^{0}$} & 0.03
(0.41, 1.04) 0.42 & 0.06 (0.20, 0.47) 0.21 \\ 
\multicolumn{1}{l}{} & \multicolumn{1}{l}{\ \ $\hat{\beta}_{2}^{X}$} & 0.00
(0.35, 0.39) 0.35 & 0.07 (0.17, 0.18) 0.18 \\ 
& \multicolumn{1}{l}{\ \ $\hat{\beta}_{2}^{\pi }$} & \textit{0.00 (0.37,
0.43) 0.37} & \textit{0.07 (0.18, 0.19) 0.19} \\ 
\multicolumn{1}{l}{(4) $\varepsilon \ $MNabs} & \multicolumn{1}{l}{$\hat{%
\beta}_{1}^{0}$} & 0.18 (0.18, 0.20) 0.26 & 0.26 (0.09, 0.10) 0.27 \\ 
\multicolumn{1}{r}{reg. false} & \multicolumn{1}{l}{\ \ $\hat{\beta}_{1}^{X}$%
} & 0.11 (0.13, 0.13) 0.17 & 0.06 (0.06, 0.06) 0.08 \\ 
& \multicolumn{1}{l}{\ \ $\hat{\beta}_{1}^{\pi }$} & \textit{0.09 (0.13,
0.14) 0.16} & \textit{0.03 (0.06, 0.07) 0.07} \\ 
& \multicolumn{1}{l}{$\hat{\beta}_{2}^{0}$} & 0.09 (0.17, 0.17) 0.19 & 0.21
(0.09, 0.09) 0.23 \\ 
& \multicolumn{1}{l}{\ \ $\hat{\beta}_{2}^{X}$} & 0.22 (0.15, 0.15) 0.27 & 
0.12 (0.07, 0.08) 0.14 \\ 
\multicolumn{1}{l}{} & \multicolumn{1}{l}{\ \ $\hat{\beta}_{2}^{\pi }$} & 
\textit{0.20 (0.15, 0.15) 0.25} & \textit{0.08 (0.07, 0.08) 0.11} \\ \hline
\multicolumn{4}{c}{SD: simulation SD; SD$_{asy}$:\ averaged asy. SD; reg.
false: MNL regression} \\ 
\multicolumn{4}{c}{misspecified; $\hat{\beta}_{d}^{0},\hat{\beta}_{d}^{X},%
\hat{\beta}_{d}^{\pi }$ with $E(Y),\ E(Y|X,D^{0d}=1),$ $E(Y|\pi
_{X}^{0d},D^{0d}=1)$} \\ \hline\hline
\end{tabular}%
$
\end{center}

\qquad In centering $Y$, $\hat{\beta}_{d}^{X}$ for $G_{X}=E(Y|X,D^{0d}=1)$
uses $1$, $(X_{0},X_{1},X_{2},X_{3})$ and their second order terms including
all interactions, whereas $\hat{\beta}_{d}^{\pi }$ for $G_{X}=E(Y|\pi
_{X}^{0d},D^{0d}=1)$ uses $1$, $(W_{1}\alpha ,W_{2}\alpha )$ and their
second order terms including all interactions. Other than for $P(D=d|X)$ and 
$(W_{1}^{\prime }\alpha ,W_{2}^{\prime }\alpha )$, the other aspects of the
simulation design are the same as for the ordered $D$ in Table 1, and the
multinomial-$D$ results are in Table 2.

\qquad As in Table 1, Panel 1 of Table 2 has the correct specification of
the $\varepsilon $ distribution and the treatment regression function; Panel
2 has the false $\varepsilon $ distribution but the correct treatment
regression function; Panel 3 has the correct $\varepsilon $ distribution but
the false treatment\ regression function; and Panel 4 has both the false $%
\varepsilon $ distribution and false regression function. In Panel 1, all
three estimators are little biased, but $\hat{\beta}_{d}^{X}$ and $\hat{\beta%
}_{d}^{\pi }$ are far more efficient than $\hat{\beta}_{d}^{0}$; $\hat{\beta}%
_{d}^{X}$ and $\hat{\beta}_{d}^{\pi }$ perform similarly. In Panel 2, $\hat{%
\beta}_{d}^{X}$ and $\hat{\beta}_{d}^{\pi }$ perform comparably but $\hat{%
\beta}_{d}^{\pi }$ is less biased than $\hat{\beta}_{d}^{X}$, and $\hat{\beta%
}_{d}^{\pi }$ and $\hat{\beta}_{d}^{X}$ dominate $\hat{\beta}_{d}^{0}$,
which is also mostly true of Panels 3 and 4.

\qquad Overall, in Table 2, $\hat{\beta}_{d}^{\pi }$ performs slightly
better than, or equally as well as, $\hat{\beta}_{d}^{X}$. Combining this
finding with the better performance of $\hat{\beta}_{d}^{\pi }$ over $\hat{%
\beta}_{d}^{X}$ in Table 1, we recommend $\hat{\beta}_{d}^{\pi }$, which is
also much easier to implement than $\hat{\beta}_{d}^{X}$.

\section{Conclusions}

\qquad In finding the effects of multiple treatments $D=0,1,...,J$, the
ubiquitous practice is constructing the dummy variables $D_{1},..,D_{J}$ to
apply the OLS of $Y$ on $(D_{1},..,D_{J},X)$ with covariates $X$. With the
potential outcomes $(Y^{0},Y^{1},...,Y^{J})$, when $\mu _{j}(X)\equiv
E(Y^{j}-Y^{0}|X)$ is not constant for some $j$, this paper showed that the
usual OLS is inconsistent, because the OLS\ $D_{d}$-slope estimates a sum of
weighted averages of \textit{all} $\mu _{1}(X),...,\mu _{J}(X)$, not just $%
\mu _{d}(X)$, which holds for any $Y$ (continuous, binary, count, ...). The
recent Covid vaccines demonstrated well how extremely treatment effects can
be heterogeneous.

\qquad The only way to prevent the \textquotedblleft
contamination\textquotedblright\ from the other treatment effects would be
isolating $D_{d}$ by using the subsample $D=0,d$ to estimate the effect of $%
D_{d}$ relative to $D_{0}$. For the subsample approach, we proposed the OLS
of $Y$ on $D_{d}-\pi _{X}^{d}$, with $\pi _{X}^{d}\equiv E(D_{d}|X,\ D=0,d)$
replaced by an estimator. Although $\pi _{X}^{d}$ can be estimated
nonparametrically, for practicality, we considered parametric approaches;
e.g., using ordered probit for ordinal $D$, and multinomial logit for
multinomial $D$. Although this would appeal to practitioners, a concern here
is possible misspecifications in the parametric estimators.

\qquad To make the subsample OLS\ robust to misspecifications in $\pi
_{X}^{d}$, we then proposed the OLS using $Y-G_{X}$ instead of $Y$, where $%
G_{X}$ is a chosen function of $X$; $G_{X}$ does not matter for the
subsample OLS consistency. We considered two forms of $G_{X}$. One is $%
E(Y|X,\ D=0,d)$ following the recent \textquotedblleft
double-debiasing\textquotedblright\ literature, and the other is $E(Y|\pi
_{X}^{0},\pi _{X}^{d},\ D=0,d)$. In our simulation study, the estimators
based on the two forms performed similarly, with the latter edging out the
former. However, $E(Y|\pi _{X}^{0},\pi _{X}^{d},\ D=0,d)$ is much easier to
estimate than $E(Y|X,\ D=0,d)$, as the former depends only on two functions
of $X$ to achieve a considerable dimension reduction compared with the
latter.

\qquad In summary, in finding multiple treatment effects, we recommend the
`robustified subsample OLS'\ of $Y-E(Y|\pi _{X}^{0},\pi _{X}^{d},\ D=0,d)$
on $D_{d}-\pi _{X}^{d}$, using only the subsample $D=0,d$. This robustified
subsample OLS is consistent for an \textquotedblleft overlap
weight\textquotedblright\ average of $\mu _{d}(\pi _{X}^{0},\pi _{X}^{d})$,
and it is a \textquotedblleft righteous way\textquotedblright\ to find the
effect of $D_{d}$, without the guilty feeling of assuming constant effects
contrary to the reality in the usual linear models.\bigskip 

\begin{center}
{\Large APPENDIX}\bigskip 
\end{center}

\textbf{Proof for Theorem 1\bigskip }

\qquad The estimands of the OLS $D_{1}$- and $D_{2}$-slopes under $\pi
_{jX}=\lambda _{jX}$ for $j=1,2$ are%
\begin{equation}
\left[ 
\begin{array}{cc}
E\{(D_{1}-\pi _{1X})^{2}\} & E\{(D_{1}-\pi _{1X})(D_{2}-\pi _{2X})\} \\ 
E\{(D_{2}-\pi _{2X})(D_{1}-\pi _{1X})\} & E\{(D_{2}-\pi _{2X})^{2}\}%
\end{array}%
\right] ^{-1}\left[ 
\begin{array}{c}
E\{(D_{1}-\pi _{1X})Y\} \\ 
E\{(D_{2}-\pi _{2X})Y\}%
\end{array}%
\right] .  \tag{A.1}
\end{equation}%
The first element of the vector involving $Y$ is%
\begin{eqnarray*}
&&E\{(D_{1}-\pi _{1X})Y\}=E[\ (D_{1}-\pi _{1X})\cdot \{\sum_{j=1}^{2}\mu
_{j}(X)D_{j}+E(Y^{0}|X)+U\}\ ] \\
&=&E\{(D_{1}-\pi _{1X})\cdot \sum_{j=1}^{2}\mu _{j}(X)D_{j}\}=E\{(D_{1}-\pi
_{1X})\cdot \sum_{j=1}^{2}\mu _{j}(X)(D_{j}-\pi _{jX})\} \\
&=&E[\sum_{j=1}^{2}E\{(D_{1}-\pi _{1X})(D_{j}-\pi _{jX})|X\}\cdot \mu
_{j}(X)]=E\{\sum_{j=1}^{2}C_{1j}(X)\cdot \mu _{j}(X)\};
\end{eqnarray*}%
the third equality holds because $\sum_{j=1}^{2}\mu _{j}(X)\pi _{jX}$ is
orthogonal to $D_{1}-\pi _{1X}$.

\qquad The determinant of the inverted matrix in (A.1) is $%
C_{11}C_{22}-C_{12}^{2}$, and (A.1) becomes%
\begin{equation*}
\frac{1}{C_{11}C_{22}-C_{12}^{2}}\left[ 
\begin{array}{cc}
C_{22} & -C_{12} \\ 
-C_{12} & C_{11}%
\end{array}%
\right] \cdot \left[ 
\begin{array}{c}
E\{C_{11}(X)\mu _{1}(X)+C_{12}(X)\mu _{2}(X)\} \\ 
E\{C_{21}(X)\mu _{1}(X)+C_{22}(X)\mu _{2}(X)\}%
\end{array}%
\right] .
\end{equation*}%
The first element of this product for the OLS $D_{1}$-slope is
\textquotedblleft $C_{11}C_{22}-C_{12}^{2}$ times\textquotedblright 
\begin{eqnarray*}
&&C_{22}\cdot E\{C_{11}(X)\mu _{1}(X)+C_{12}(X)\mu _{2}(X)\}-C_{12}\cdot
E\{C_{21}(X)\mu _{1}(X)+C_{22}(X)\mu _{2}(X)\} \\
&=&E\{\ C_{22}C_{11}(X)\mu _{1}(X)+C_{22}C_{12}(X)\mu
_{2}(X)-C_{12}C_{21}(X)\mu _{1}(X)-C_{12}C_{22}(X)\mu _{2}(X)\ \} \\
&=&E[\ \{C_{22}C_{11}(X)-C_{12}C_{21}(X)\}\cdot \mu
_{1}(X)+\{C_{22}C_{12}(X)-C_{12}C_{22}(X)\}\cdot \mu _{2}(X)\ ].
\end{eqnarray*}%
Dividing this by $C_{11}C_{22}-C_{12}^{2}$ yields $E\{\omega _{11}(X)\mu
_{1}(X)+\omega _{12}(X)\mu _{2}(X)\}$ in (2.5), and then switching $1$ and $%
2 $ for the OLS $D_{2}$-slope renders $E\{\omega _{22}(X)\mu _{2}(X)+\omega
_{21}(X)\mu _{1}(X)\}$. Take $E(\cdot )$ on the weights to see (2.6).\bigskip

\textbf{Proof for Theorem 1 Extended to Four Categories\bigskip }

\qquad The estimand of the OLS is%
\begin{equation}
\left[ 
\begin{array}{ccc}
C_{11} & C_{12} & C_{13} \\ 
C_{21} & C_{22} & C_{23} \\ 
C_{31} & C_{32} & C_{33}%
\end{array}%
\right] ^{-1}\left[ 
\begin{array}{c}
E\{\sum_{j=1}^{3}C_{1j}(X)\mu _{j}(X)\} \\ 
E\{\sum_{j=1}^{3}C_{2j}(X)\mu _{j}(X)\} \\ 
E\{\sum_{j=1}^{3}C_{3j}(X)\mu _{j}(X)\}%
\end{array}%
\right] .  \tag{A.2}
\end{equation}%
For the first matrix, to find the inverse that is the adjoint divided by the
determinant, note that the determinant (assumed to be non-zero) and the
adjoint are%
\begin{eqnarray*}
&&Det\equiv
C_{11}(C_{22}C_{33}-C_{23}C_{32})-C_{21}(C_{12}C_{33}-C_{13}C_{32})+C_{31}(C_{12}C_{23}-C_{13}C_{22}),
\\
&&\left[ 
\begin{array}{ccc}
C_{22}C_{33}-C_{23}C_{32} & -(C_{12}C_{33}-C_{13}C_{32}) & 
C_{12}C_{23}-C_{13}C_{22} \\ 
-(C_{21}C_{33}-C_{23}C_{31}) & C_{11}C_{33}-C_{13}C_{31} & 
-(C_{11}C_{23}-C_{13}C_{21}) \\ 
C_{21}C_{32}-C_{22}C_{31} & -(C_{11}C_{32}-C_{12}C_{31}) & 
C_{11}C_{22}-C_{12}C_{21}%
\end{array}%
\right] .
\end{eqnarray*}

\qquad The first element of the OLS estimand is the product of the first row
of the adjoint times the last vector in (A.2) divided by the determinant:%
\begin{eqnarray*}
&&[\ (C_{22}C_{33}-C_{23}C_{32})E\{\sum_{j=1}^{3}C_{1j}(X)\mu
_{j}(X)\}-(C_{12}C_{33}-C_{13}C_{32})E\{\sum_{j=1}^{3}C_{2j}(X)\mu _{j}(X)\}
\\
&&+(C_{12}C_{23}-C_{13}C_{22})E\{\sum_{j=1}^{3}C_{3j}(X)\mu _{j}(X)\}\ ]/Det.
\end{eqnarray*}%
Collecting the terms with $\mu _{1}(X)$, $\mu _{2}(X)$, $\mu _{3}(X)$ yields 
$E\{\omega _{11}(X)\mu _{1}(X)\}$, $E\{\omega _{12}(X)\mu _{2}(X)\}$, $%
E\{\omega _{13}(X)\mu _{3}(X)\}$ respectively, where, writing $C_{jd}(X)$
just as $C_{jdX}$,%
\begin{eqnarray*}
\omega _{11}(X) &\equiv &\frac{%
(C_{22}C_{33}-C_{23}C_{32})C_{11X}-(C_{12}C_{33}-C_{13}C_{32})C_{21X}+(C_{12}C_{23}-C_{13}C_{22})C_{31X}%
}{Det}, \\
\omega _{12}(X) &\equiv &\frac{%
(C_{22}C_{33}-C_{23}C_{32})C_{12X}-(C_{12}C_{33}-C_{13}C_{32})C_{22X}+(C_{12}C_{23}-C_{13}C_{22})C_{32X}%
}{Det}, \\
\omega _{13}(X) &\equiv &\frac{%
(C_{22}C_{33}-C_{23}C_{32})C_{13X}-(C_{12}C_{33}-C_{13}C_{32})C_{23X}+(C_{12}C_{23}-C_{13}C_{22})C_{33X}%
}{Det}.
\end{eqnarray*}

\qquad Therefore, $\sum_{j=1}^{3}E\{\omega _{1j}(X)\mu _{j}(X)\}$ is the
estimand of the OLS $D_{1}$-slope. Since the expected value of the numerator
of $\omega _{11}(X)$ is the same as $Det$, $E\{\omega _{11}(X)\}=1$ holds.
In contrast, the expected values of the numerators of $\omega _{12}(X)$ and $%
\omega _{13}(X)$ are zero due to all terms cancelled to render $E\{\omega
_{12}(X)\}=0$ and $E\{\omega _{13}(X)\}=0$. The estimands of the $D_{2}$ and 
$D_{3}$ slopes can be analogously obtained.\bigskip

\textbf{Proof for Theorem 2\bigskip }

\qquad Note $D^{0d}D_{d}=D_{d}$, $E(Y|X,D^{0d}=1)=E(D^{0d}Y|X)/E(D^{0d}|X)$,
and%
\begin{eqnarray}
&&D^{0d}Y=D^{0d}\{\mu _{d}(X)D_{d}+E(Y^{0}|X)+U\}  \TCItag{A.3} \\
&\Longrightarrow &E(D^{0d}Y|X)=\mu _{d}(X)E(D_{d}|X)+E(Y^{0}|X)E(D^{0d}|X)%
\text{ \ \ \ \ (taking }E(\cdot |X)\text{)}  \notag \\
&\Longrightarrow &D^{0d}\frac{E(D^{0d}Y|X)}{E(D^{0d}|X)}=D^{0d}\mu _{d}(X)%
\frac{E(D_{d}|X)}{E(D^{0d}|X)}+D^{0d}E(Y^{0}|X)\text{ (multiplying }\frac{%
D^{0d}}{E(D^{0d}|X)}\text{)}  \notag \\
&\Longrightarrow &D^{0d}E(Y|X,D^{0d}=1)=D^{0d}\mu _{d}(X)\pi
_{X}^{d}+D^{0d}E(Y^{0}|X)  \TCItag{A.4} \\
&\Longrightarrow &D^{0d}\{Y-E(Y|X,D^{0d}=1)\}=\mu _{d}(X)D^{0d}(D_{d}-\pi
_{X}^{d})+D^{0d}U,  \TCItag{A.5}
\end{eqnarray}%
subtracting (A.4) from (A.3).

\qquad For the estimand of the OLS to (A.5), $E(Y|X,D^{0d}=1)$ can be
ignored, because%
\begin{equation}
\frac{E[\ D^{0d}(D_{d}-\pi _{X}^{d})\cdot \{Y-E(Y|X,D^{0d}=1)\}\ ]}{E\{\
D^{0d}(D_{d}-\pi _{X}^{d})^{2}\ \}}=\frac{E\{D^{0d}(D_{d}-\pi _{X}^{d})\cdot
D^{0d}Y\}}{E\{D^{0d}(D_{d}-\pi _{X}^{d})^{2}\}}  \tag{A.6}
\end{equation}%
where $E(Y|X,D^{0d}=1)$ is orthogonal to $D^{0d}(D_{d}-\pi _{X}^{d})$ as
(3.3) shows. Since $E(Y^{0}|X)$ is also orthogonal to $D^{0d}(D_{d}-\pi
_{X}^{d})$, substituting (A.3) into (A.6) renders%
\begin{equation}
\frac{E\{\ D^{0d}(D_{d}-\pi _{X}^{d})\cdot \mu _{d}(X)D_{d}\ \}}{%
E\{D^{0d}(D_{d}-\pi _{X}^{d})^{2}\}}=\frac{E[\ E\{D^{0d}(D_{d}-\pi
_{X}^{d})D_{d}|X\}\cdot \mu _{d}(X)\ ]}{E\{D^{0d}(D_{d}-\pi _{X}^{d})^{2}\}}.
\tag{A.7}
\end{equation}

\qquad For the denominator of (A.7), observe%
\begin{eqnarray}
&&E\{D^{0d}(D_{d}-\pi _{X}^{d})^{2}\}=E[\ D^{0d}\{D_{d}-2D_{d}\pi
_{X}^{d}+(\pi _{X}^{d})^{2}\}\ ]  \notag \\
&=&E\{D_{d}-2D_{d}\pi _{X}^{d}+D^{0d}(\pi
_{X}^{d})^{2}\}=E\{E(D_{d}|X)-2E(D_{d}|X)\pi _{X}^{d}+E(D^{0d}|X)(\pi
_{X}^{d})^{2}\}  \notag \\
&=&E[E(D_{d}|X)-2\frac{\{E(D_{d}|X)\}^{2}}{E(D^{0d}|X)}+\frac{%
\{E(D_{d}|X)\}^{2}}{E(D^{0d}|X)}]=E[E(D_{d}|X)-\frac{\{E(D_{d}|X)\}^{2}}{%
E(D^{0d}|X)}]  \notag \\
&=&E\{\pi _{X}^{d}E(D^{0d}|X)-(\pi _{X}^{d})^{2}E(D^{0d}|X)\}=E\{\pi
_{X}^{d}(1-\pi _{X}^{d})E(D^{0d}|X)\}.  \TCItag{A.8}
\end{eqnarray}%
As for the numerator of (A.7), since $E\{D^{0d}(D_{d}-\pi _{X}^{d})|X\}=\pi
_{dX}-\pi _{dX}=0$,%
\begin{eqnarray*}
&&E\{D^{0d}(D_{d}-\pi _{X}^{d})D_{d}|X\}=E\{D^{0d}(D_{d}-\pi
_{X}^{d})D_{d}|X\}-E\{D^{0d}(D_{d}-\pi _{X}^{d})|X\}\pi _{X}^{d} \\
&&\ =E\{D^{0d}(D_{d}-\pi _{X}^{d})^{2}|X\}=\pi _{X}^{d}(1-\pi
_{X}^{d})E(D^{0d}|X)\text{ \ \ \ \ (in view of (A.8)).}
\end{eqnarray*}%
This and (A.8) yield $E\{\omega _{ow}^{d}(X)\mu _{d}(X)\}$.\bigskip

\textbf{Proof for Theorem 3\medskip }

\qquad Recalling $\pi _{X}^{0d}\equiv (\pi _{0X},\pi _{dX})^{\prime }$, note
that $D_{j}\amalg (Y^{0},Y^{d})|X$ implies $D_{j}\amalg (Y^{0},Y^{d})|\pi
_{X}^{0d}$ for $j=0,d$, due to the binary nature of $D_{j}$ and the first
and last expressions in%
\begin{eqnarray}
&&E(D_{j}|Y^{0},Y^{d},\pi
_{X}^{0d})=E\{E(D_{j}|Y^{0},Y^{d},X)|Y^{0},Y^{d},\pi
_{X}^{0d}\}=E\{E(D_{j}|X)|Y^{0},Y^{d},\pi _{X}^{0d}\}  \notag \\
&&\ =E(D_{j}|X)=E(D_{j}|\pi _{X}^{0d});  \TCItag{A.9}
\end{eqnarray}%
the last equality holds by taking $E(\cdot |\pi _{X}^{0d})$ on $\pi
_{jX}\equiv E(D_{j}|X)$, $j=0,d$. Replacing $D_{j}$ with $D^{0d}=D_{0}+D_{d}$
in (A.9) further yields $D^{0d}\amalg (Y^{0},Y^{d})|\pi _{X}^{0d}$. Using
this,%
\begin{eqnarray*}
&&E(Y|\pi _{X}^{0d},D^{0d}=1)=E(Y^{d}-Y^{0}|\pi
_{X}^{0d},D^{0d}=1)D_{d}+E(Y^{0}|\pi _{X}^{0d},D^{0d}=1) \\
&&\ =E(Y^{d}-Y^{0}|\pi _{X}^{0d})D_{d}+E(Y^{0}|\pi _{X}^{0d})=\mu _{d}(\pi
_{X}^{0d})D_{d}+E(Y^{0}|\pi _{X}^{0d}).
\end{eqnarray*}%
The first and last expressions here along with $U^{\prime }\equiv Y-E(Y|\pi
_{X}^{0d},D^{0d}=1)$ render%
\begin{eqnarray}
&&Y=\mu _{d}(\pi _{X}^{0d})D_{d}+E(Y^{0}|\pi _{X}^{0d})+U^{\prime },\ \ \ \
\ E(U|\pi _{X}^{0d},D^{0d}=1)=0  \notag \\
&\Longrightarrow &D^{0d}Y=D^{0d}\mu _{d}(\pi
_{X}^{0d})D_{d}+D^{0d}E(Y^{0}|\pi _{X}^{0d})+D^{0d}U^{\prime }\text{
(multiplying }D^{0d}\text{).}  \TCItag{A.10}
\end{eqnarray}

\qquad Take $E(\cdot |\pi _{X}^{0d})$ on this $D^{0d}Y$ equation, and then
multiply by $D^{0d}/E(D^{0d}|\pi _{X}^{0d})$:%
\begin{eqnarray}
&&E(D^{0d}Y|\pi _{X}^{0d})=\mu _{d}(\pi _{X}^{0d})E(D_{d}|\pi
_{X}^{0d})+E(Y^{0}|\pi _{X}^{0d})E(D^{0d}|\pi _{X}^{0d})  \notag \\
&&\ \Longrightarrow D^{0d}\frac{E(D^{0d}Y|\pi _{X}^{0d})}{E(D^{0d}|\pi
_{X}^{0d})}=D^{0d}\mu _{d}(\pi _{X}^{0d})\frac{E(D_{d}|\pi _{X}^{0d})}{%
E(D^{0d}|\pi _{X}^{0d})}+D^{0d}E(Y^{0}|\pi _{X}^{0d})  \notag \\
&&\ \Longrightarrow D^{0d}E(Y|\pi _{X}^{0d},D^{0d}=1)=D^{0d}\mu _{d}(\pi
_{X}^{0d})\pi _{X}^{d}+D^{0d}E(Y^{0}|\pi _{X}^{0d})\text{ \ \ because} 
\TCItag{A.11} \\
&&\frac{E(D^{0d}Y|\pi _{X}^{0d})}{E(D^{0d}|\pi _{X}^{0d})}=E(Y|\pi
_{X}^{0d},D^{0d}=1),\text{\ \ \ }\frac{E(D_{d}|\pi _{X}^{0d})}{E(D^{0d}|\pi
_{X}^{0d})}=\frac{E(D_{d}|X)}{E(D^{0d}|X)}=\pi _{X}^{d}\text{ \ due to (A.9)}%
.  \notag
\end{eqnarray}%
Finally, subtracting (A.11) from (A.10) yields the subsample $Y$ equation:%
\begin{equation}
D^{0d}\{Y-E(Y|\pi _{X}^{0d},D^{0d}=1)\}=\mu _{d}(\pi
_{X}^{0d})D^{0d}(D_{d}-\pi _{X}^{d})+D^{0d}U^{\prime }.  \tag{A.12}
\end{equation}

\qquad The OLS\ estimand of $D^{0d}\{Y-E(Y|\pi _{X}^{0d},D^{0d}=1)\}$ on $%
D^{0d}(D-\pi _{X}^{d})$ is, using (A.12),%
\begin{eqnarray}
&&\frac{E[\ D^{0d}(D_{d}-\pi _{X}^{d})\cdot D^{0d}\{Y-E(Y|\pi
_{X}^{0d},D^{0d}=1)\}\ ]}{E\{\ D^{0d}(D_{d}-\pi _{X}^{d})^{2}\ \}}  \notag \\
&&\ =\frac{E\{D^{0d}(D_{d}-\pi _{X}^{d})^{2}\cdot \mu _{d}(\pi _{X}^{0d})\}}{%
E\{D^{0d}(D_{d}-\pi _{X}^{d})^{2}\}}+\frac{E\{D^{0d}(D_{d}-\pi
_{X}^{d})U^{\prime }\}}{E\{D^{0d}(D_{d}-\pi _{X}^{d})^{2}\}}.  \TCItag{A.13}
\end{eqnarray}%
The first term in (A.13) renders $\beta _{d}$ in Theorem 2, because $\omega
_{ow}^{d}(X)$ in (2.8) depends on $X$ only through $\pi _{X}^{0d}$,
recalling $E(D^{0d}|X)=E(D^{0d}|\pi _{X}^{0d})$: the same estimand is
estimated using $\pi _{X}^{0d}$ instead of $X$. The following proves that
the second term of (A.13) is zero.

\qquad Observe $E\{D^{0d}(D_{d}-\pi _{X}^{d})U^{\prime }\}=E(D_{d}U^{\prime
})-E(D^{0d}\pi _{X}^{d}U^{\prime })$, which is in turn equal to%
\begin{eqnarray}
&&E\{E(D_{d}U^{\prime }|\pi _{X}^{0d})\}-E\{E(D^{0d}U^{\prime }|\pi
_{X}^{0d})\pi _{X}^{d}\}  \notag \\
&=&E\{E(U^{\prime }|\pi _{X}^{0d},D_{d}=1)P(D_{d}=1|\pi
_{X}^{0d})\}-E\{E(D^{0d}U^{\prime }|\pi _{X}^{0d})\pi _{X}^{d}\}. 
\TCItag{A.14}
\end{eqnarray}%
The second term of (A.14) is zero because, due to $U^{\prime }\equiv
Y-E(Y|\pi _{X}^{0d},D^{0d}=1)$,%
\begin{equation*}
E\{E(D^{0d}U^{\prime }|\pi _{X}^{0d})\pi _{X}^{d}\}=E\{\ E(U^{\prime }|\pi
_{X}^{0d},D^{0d}=1)\cdot P(D^{0d}=1|\pi _{X}^{0d})\pi _{X}^{d}\ \}=0.
\end{equation*}%
The first term is also zero, because $D^{0d}=1$ is \textquotedblleft
finer\textquotedblright\ than $D_{d}=1\Longleftrightarrow (D_{0}=0,D_{d}=1)$:%
\begin{equation*}
E(U^{\prime }|\pi _{X}^{0d},D_{d}=1)=E\{E(U^{\prime }|\pi
_{X}^{0d},D^{0d}=1)|\pi _{X}^{0d},D_{0}=0,D_{d}=1\}=0.
\end{equation*}%
In words, $E(U^{\prime }|\pi _{X}^{0d},D^{0d}=1)=0$ holds for all possible
values of $(D_{0},D_{d})$ such that $D^{0d}=1$, and consequently, $%
E(U^{\prime }|\pi _{X}^{0d},D_{0}=0,D_{d}=1)=0$ also holds.\bigskip

\textbf{Proof for Moment-Derivatives of Robustified Subsample OLS}$%
_{psr}\medskip $

\qquad The moment condition for the robustified subsample OLS$_{psr}$ is%
\begin{equation*}
E[\ D^{0d}\{Y-E(Y|\pi _{0X},\pi _{dX},D^{0d}=1)-\beta _{d}(D-\pi
_{X}^{d})\}\cdot (D-\pi _{X}^{d})\ ]=0.
\end{equation*}%
Replace $\pi _{0X}$ with $\pi _{0X}+ap_{X}$, $\pi _{dX}$ with $\pi
_{dX}+bq_{X}$, and $E(Y|\pi _{0X},\pi _{dX},D^{0d}=1)$ with $E(Y|\pi
_{0X}+ap_{X},\pi _{dX}+bq_{X},D^{0d}=1)+cL(\pi _{0X}+ap_{X},\pi
_{dX}+bq_{X}) $ for a function $L(\pi _{0X},\pi _{dX})$:%
\begin{eqnarray*}
&&E[\ D^{0d}\{Y-E(Y|\pi _{0X}+ap_{X},\pi _{dX}+bq_{X},D^{0d}=1)-cL(\pi
_{0X}+ap_{X},\pi _{dX}+bq_{X}) \\
&&\ \ \ -\beta _{d}(D-\frac{\pi _{dX}+bq_{X}}{\pi _{0X}+ap_{X}+\pi
_{dX}+bq_{X}})\}\cdot (D-\frac{\pi _{dX}+bq_{X}}{\pi _{0X}+ap_{X}+\pi
_{dX}+bq_{X}})\ ].
\end{eqnarray*}

\qquad Denoting functions of $\pi _{X}^{0d}\equiv (\pi _{0X},\pi
_{dX})^{\prime }$ just as $H_{1}(\pi _{X}^{0d})$, $H_{2}(\pi _{X}^{0d})$ and
so on, note, analogously to (3.3),%
\begin{equation}
E\{H_{1}(\pi _{X}^{0d})D^{0d}(D_{d}-\pi _{X}^{d})\}=E[H_{1}(\pi
_{X}^{0d})\cdot E\{D^{0d}(D_{d}-\pi _{X}^{d})|X\}]=0.  \tag{A.15}
\end{equation}%
Also, because $E(D^{0d}Y|\pi _{X}^{0d})=E(D^{0d}|\pi _{X}^{0d})E(Y|\pi
_{X}^{0d})$ due to $D_{j}\amalg (Y^{0},Y^{d})|\pi _{X}^{0d}$ for $j=0,d$ ($Y$
in $D^{0d}Y$ is either $Y^{0}$ and $Y^{d}$),%
\begin{eqnarray}
&&E[H_{1}(\pi _{X}^{0d})D^{0d}\{Y-E(Y|\pi _{X}^{0d})\}]=E(\ H_{1}(\pi
_{X}^{0d})E[D^{0d}\{Y-E(Y|\pi _{X}^{0d})\}|\pi _{X}^{0d}]\ )  \notag \\
&&\ =E[\ H_{1}(\pi _{X}^{0d})\{E(D^{0d}Y|\pi _{X}^{0d})-E(D^{0d}|\pi
_{X}^{0d})E(Y|\pi _{X}^{0d})\}\ ]=0.  \TCItag{A.16}
\end{eqnarray}

\qquad Differentiate the above moment wrt $a$ at $(a,b,c)=(0,0,0)$: using
(A.15) and (A.16),%
\begin{eqnarray*}
&&E[\ \{-\frac{\partial E(Y|\pi _{0X}+ap_{X},\pi _{dX}+bq_{X},D^{0d}=1)}{%
\partial a}|_{a,b,c=0}+\beta _{d}H_{1}(\pi _{X}^{0d})\}p_{X}D^{0d}(D-\pi
_{X}^{d})\ ] \\
&&\ -D^{0d}\{Y-E(Y|\pi _{0X},\pi _{dX},D^{0d}=1)-\beta _{d}(D-\pi
_{X}^{d})\}p_{X}\cdot H_{2}(\pi _{X}^{0d})\ ] \\
&=&-E[\ D^{0d}\{Y-E(Y|\pi _{0X},\pi _{dX},D^{0d}=1)\}p_{X}\cdot H_{2}(\pi
_{X}^{0d})\ ].
\end{eqnarray*}%
This is not zero in general, but close to zero to the extent that $p_{X}$ is
well approximated by power functions of $(\pi _{0X},\pi _{dX})$. Doing
analogously, we can make the same statement for the derivative wrt $b$ at $%
(a,b,c)=(0,0,0)$. As for the derivative wrt $c$ at $(a,b,c)=(0,0,0)$, it is $%
E\{-L(\pi _{X}^{0d})\cdot D^{0d}(D-\pi _{X}^{d})\}=0$: the robustified
subsample OLS\ is locally robust at least to the misspecified $E(Y|\pi
_{0X},\pi _{dX},D^{0d}=1)$.\bigskip

\textbf{Proof for Theorem 4\medskip }

\qquad Writing $m(\hat{\beta}_{d}^{q},\hat{\alpha},\hat{\gamma}%
^{d};X_{i},Y_{i})$ in (3.5) just as $m(\hat{\beta}_{d}^{q},\hat{\alpha},\hat{%
\gamma}^{d})$, expand this around $\beta _{d}$:%
\begin{equation*}
0=\frac{1}{\sqrt{N}}\sum_{i}m(\beta _{d},\hat{\alpha},\hat{\gamma}^{d})+%
\frac{1}{N}\sum_{i}\frac{\partial m(\hat{\beta}_{d}^{q\ast },\hat{\alpha},%
\hat{\gamma}^{d})}{\partial b}\sqrt{N}(\hat{\beta}_{d}^{q}-\beta _{d})\text{
\ \ for a }\hat{\beta}_{d}^{q\ast }\in (\hat{\beta}_{d}^{q},\beta _{d}).
\end{equation*}%
Letting `$E^{-1}(\cdot )$' be $\{E(\cdot )\}^{-1}$, write this as, with $g$
for $\gamma ^{d}\equiv (\gamma _{0}^{d},...,\gamma _{q}^{d})^{\prime }$,%
\begin{eqnarray}
&&\sqrt{N}(\hat{\beta}_{d}^{q}-\beta _{d})=-E^{-1}\{\frac{\partial m(\beta
_{d},\alpha ,\gamma ^{d})}{\partial b}\}\cdot \lbrack \frac{1}{\sqrt{N}}%
\sum_{i}m(\beta _{d},\alpha ,\gamma ^{d})  \TCItag{A.17} \\
&&+E\{\frac{\partial m(\beta _{d},\alpha ,\gamma ^{d})}{\partial a^{\prime }}%
\}\sqrt{N}(\hat{\alpha}-\alpha )+E\{\frac{\partial m(\beta _{d},\alpha
,\gamma ^{d})}{\partial g^{\prime }}\}\sqrt{N}(\hat{\gamma}^{d}-\gamma
^{d})]+o_{p}(1).  \notag
\end{eqnarray}

\qquad Note%
\begin{equation*}
\frac{\partial m(\beta _{d},\alpha ,\gamma ^{d})}{\partial b}=-D^{0d}\{D_{d}-%
\frac{P_{d}(\alpha ;X)}{P_{0}(\alpha ;X)+P_{d}(\alpha ;X)}\}^{2}
\end{equation*}%
which yields $(N^{-1}\sum_{i}D_{i}^{0d}\hat{\varepsilon}_{di}^{2})^{-2}$ in
the variance. As for $N^{-1}\sum_{i}D_{i}^{0d}(\hat{V}_{di}\hat{\varepsilon}%
_{di}+\hat{L}_{d}\hat{\eta}_{i})^{2}$, observe%
\begin{equation*}
\frac{\partial m(\beta _{d},\alpha ,\gamma ^{d})}{\partial g^{\prime }}%
=-D^{0d}\{D_{d}-\frac{P_{d}(\alpha ;X)}{P_{0}(\alpha ;X)+P_{d}(\alpha ;X)}%
\}\{1,(X^{\prime }\kappa ),...,(X^{\prime }\kappa )^{q}\}.
\end{equation*}%
$E\{\partial m(\beta _{d},\alpha ,\gamma ^{d})/\partial g^{\prime }\}=0$ due
to $E\{D^{0d}(D_{d}-\pi _{X}^{d})|X\}=0$. Letting $\eta $ be influence
functions for $\hat{\alpha}$, (A.17) becomes the following to yield the
variance in Theorem 4:%
\begin{eqnarray*}
&&\sqrt{N}(\hat{\beta}_{d}^{q}-\beta _{d})=-E^{-1}\{\frac{\partial m(\beta
_{d},\alpha ,\gamma ^{d})}{\partial b^{\prime }}\}\cdot \frac{1}{\sqrt{N}}%
\sum_{i}\zeta _{di}+o_{p}(1), \\
&&\zeta _{di}\equiv m(\beta _{d},\alpha ,\gamma ^{d})+E\{\frac{\partial
m(\beta _{d},\alpha ,\gamma ^{d})}{\partial a^{\prime }}\}\eta
_{i}=D_{i}^{0d}V_{di}\varepsilon _{di}+L_{d}\eta _{i},\ L_{d}\equiv E\{\frac{%
\partial m(\beta _{d},\alpha ,\gamma ^{d})}{\partial a^{\prime }}\}.
\end{eqnarray*}%
For the actual implementation, we recommend using numerical derivatives for $%
L_{d}$.\bigskip 

\begin{center}
{\LARGE REFERENCES}
\end{center}

\qquad Angrist, J.D., 1998, Estimating the labor market impact of voluntary
military service using social security data on military applicants,
Econometrica 66, 249-288.

\qquad Angrist, J.D. and J.S. Pischke, 2009, Mostly harmless econometrics,
Princeton University Press.

\qquad Athey, S., J. Tibshirani and S. Wager, 2019, Generalized random
forests, Annals of Statistics 47, 1148-1178.

\qquad Cheng. C., F. Li, L.E. Thomas and F. Li, 2022, Addressing extreme
propensity scores in estimating counterfactual survival functions via the
overlap weights, American Journal of Epidemiology 191, 1140-1151.

\qquad Chernozhukov, V., D. Chetverikov, M. Demirer, E. Duflo, C. Hansen, W.
Newey and J. Robins, 2018, Double/debiased machine learning for treatment
and structural parameters, Econometrics Journal 21, C1-C68.

\qquad Chernozhukov, V., J.C. Escanciano, H. Ichimura, W.K. Newey and J.M.
Robins, 2022, Locally robust semiparametric estimation, Econometrica 90,
1501-1535.

\qquad Choi, J.Y. and M.J. Lee, 2023, Overlap weight and propensity score
residual for heterogeneous effects:\ a review with extensions, Journal of
Statistical Planning and Inference 222, 22-37.

\qquad Imbens, G.W., 2000, The role of the propensity score in estimating
dose-response functions, Biometrika 87, 706-710.

\qquad Lee, M.J., 2010, Micro-econometrics: methods of moments and limited
dependent variables, Springer.

\qquad Lee, M.J. 2018, Simple least squares estimator for treatment effects
using propensity score residuals, Biometrika 105, 149-164.

\qquad Lee, M.J., 2021, Instrument residual estimator for any response
variable with endogenous binary treatment, Journal of the Royal Statistical
Society (Series B) 83, 612-635.

\qquad Lee, M.J., G. Lee and J.Y. Choi, 2023,\ Linear probability model
revisited: why it works and how it should be specified, Sociological Methods
\& Research, forthcoming.

\qquad Lee, M.J. and S.H. Lee, 2022, Review and comparison of treatment
effect estimators using propensity and prognostic scores, International
Journal of Biostatistics 18, 357-380.

\textbf{\qquad }Li, L. and T. Greene, 2013, A weighting analogue to pair
matching in propensity score analysis, International Journal of
Biostatistics 9, 215-234.

\qquad Li, F. and F. Li, 2019, Propensity score weighting for causal
inference with multiple treatments, Annals of Applied Statistics 13,
2389-2415.

\qquad Li, F., K.L. Morgan and A.M. Zaslavsky, 2018, Balancing covariates
via propensity score weighting, Journal of the American Statistical
Association 113, 390-400.

\qquad Li, F., L.E. Thomas and F. Li, 2019, Addressing extreme propensity
scores via the overlap weights, American Journal of Epidemiology 188,
250-257.

\qquad Mao, H. and L. Li, 2020, Flexible regression approach to propensity
score analysis and its relationship with matching and weighting, Statistics
in Medicine 39, 2017-2034.

\qquad Mao, H., L. Li and T. Greene, 2019, Propensity score weighting
analysis and treatment effect discovery, Statistical Methods in Medical
Research 28, 2439-2454.

\qquad Mao, H., L. Li, W. Yang and Y. Shen, 2018, On the propensity score
weighting analysis with survival outcome: estimands, estimation, and
inference. Statistics in Medicine 37, 3745-3763.

\qquad Nie, X. and S. Wager, 2021, Quasi-oracle estimation of heterogeneous
treatment effects, Biometrika 108, 299-319.

\qquad Robins, J.M., S.D. Mark and W.K. Newey, 1992, Estimating exposure
effects by modelling the expectation of exposure conditional on confounders,
Biometrics 48, 479-495.

\qquad Robinson, P.M., 1988, Root-N consistent semiparametric regression,
Econometrica 56, 931-954.

\qquad Thomas L.E., F. Li and M.J. Pencina, 2020, Overlap weighting: a
propensity score method that mimics attributes of a randomized clinical
trial, Journal of the American Medical Association, 323, 2417-2418.

\qquad Vansteelandt, S. and R.M. Daniel, 2014, On regression adjustment for
the propensity score, Statistics in Medicine 33, 4053-4072.

\end{document}